\documentclass[superscriptaddress,aps,onecolumn]{revtex4}
\usepackage{multirow,enumerate,epsfig,graphics,amssymb,amsmath,subeqnarray,mathrsfs,color,xcolor}
\usepackage{color,epsfig,graphics,amssymb,amsmath,subeqnarray,graphicx,amsthm,subfigure,mathrsfs}

\usepackage{epstopdf}

\definecolor{cream}{RGB}{222,217,201}

\newcommand{\dd}{\mathrm{d}}
\newcommand{\ee}{\mathrm{e}}

\newcommand{\Fb}{\mathbf{F}}

\newcommand{\nb}{\mathbf{n}}
\newcommand{\Rb}{\mathbf{R}}

\newcommand{\Omegab}{\boldsymbol{\Omega}}
\newcommand{\sigmab}{\boldsymbol{\sigma}}

\newcommand{\ub}{\mathbf{u}}
\newcommand{\eb}{\mathbf{e}}

\newcommand{\Ub}{\mathbf{U}}\newcommand{\xb}{\mathbf{x}}\newcommand{\rb}{\mathbf{r}}

\newcommand{\Pe}{\mbox{Pe}}

\let\grad\nabla
\let\grad\nabla

\newcommand{\pard}[2]{\frac{\partial #1}{\partial #2}}\newcommand{\totd}[2]{\frac{\mathrm{d}#1}{\mathrm{d}#2}}

\newcommand{\Rey}{\mbox{Re}}

\makeatletter
\def\sgn{\mathop{\operator@font sgn}}
\def\threevdots{\vbox{\baselineskip1\p@ \lineskiplimit\z@
  \kern6\p@\hbox{.}\hbox{.}\hbox{.}}}
\makeatother

\begin{document}
\title{Clustering-induced self-propulsion of isotropic autophoretic particles}
\author{Akhil Varma}
\email{akhil.varma@ladhyx.polytechnique.fr}
\affiliation{LadHyX -- D\'epartement de M\'ecanique, Ecole Polytechnique -- CNRS, 91128 Palaiseau, France.}
\author{Thomas D. Montenegro-Johnson}
\email{t.d.johnson@bham.ac.uk}
\affiliation{School of Mathematics, University of Birmingham, Edgbaston, Birmingham B15 2TT, United Kingdom}
\author{S\'ebastien Michelin}
\email{sebastien.michelin@ladhyx.polytechnique.fr}
\affiliation{LadHyX -- D\'epartement de M\'ecanique, Ecole Polytechnique -- CNRS, 91128 Palaiseau, France.}
\date{\today}

\begin{abstract}
Self-diffusiophoretic particles exploit local concentration gradients of a solute species in order to self-propel at the micron scale. While an isolated chemically- and geometrically-isotropic particle cannot swim, we show that it can achieve self-propulsion through interactions with other individually-non-motile particles by forming geometrically-anisotropic clusters via phoretic and hydrodynamic interactions. This result identifies a new route to symmetry-breaking for the concentration field and to self-propulsion, that is not based on an anisotropic design, but on the collective dynamics of identical and homogeneous active particles.  Using full numerical simulations as well as theoretical modelling of the clustering process, the statistics of the propulsion properties are obtained for arbitrary initial arrangement of the particles. The robustness of these results to thermal noise, and more generally the effect of Brownian motion of the particles, is also discussed.

\end{abstract}
\maketitle

\section{Introduction}
\label{sec:intro}

Achieving self-propulsion through fluids at the micron scale entails many challenges associated with the dynamics of the surrounding fluid medium, including overcoming the dominant effect of viscous dissipation that precludes any inertia-related history effect, and breaking temporal and spatial symmetries in the flow forcing~\cite{purcell77,happelbrenner}. Swimming micro-organisms  (e.g. bacteria or swimming algae) achieve self-propulsion using the propagation of waves along their cilia or flagella, in order to generate periodic yet non-reciprocal strokes~\cite{brennen1977,lauga2009,lauga2016}. These biological systems have inspired the experimental design of many artificial swimmers in the lab~\cite{najafi04,grosjean16,ghosh09,dreyfus05}. However, such mechanical systems present fundamental and practical limitations, such as moving parts or their reliance on an external, often macroscopic, actuation (e.g. unsteady magnetic field).

Catalytic colloids represent another category of artificial micro-swimmers that rely instead on self-generated gradients in the physico-chemical properties of their immediate environment~\cite{moran2017}. Self-diffusiophoresis is such an example, where a chemically-active colloid moves in response to a solute concentration gradient that is either produced or consumed at its surface through a catalytic reaction~\cite{julicher2009,brady2011}. Other phoretic systems rely on self-generated gradients of temperature (thermophoresis) or electric potential (electrophoresis) to self-propel~\cite{yadav2015,duan2015}. Such autophoretic motion fundamentally requires two distinct physico-chemical properties of the particle. Short-range interactions between its surface and the surrounding solute generate an effective slip motion of the fluid in response to gradients in surface concentration, a property generally termed as phoretic \textit{mobility}, $\mathcal{M}$~\cite{anderson1989}. The second property, the chemical \emph{activity}, $\mathcal{A}$, refers to the particle's ability to absorb or release solute through chemical reactions at their surface \cite{Golestanian07}. 

A major advantage of such active colloids in comparison with macroscopically-actuated microswimmers lies in the local nature of the interaction phenomena at the heart of their propulsion, which allows for complex collective behaviour; in contrast, collective motion of magnetic swimmers is essentially driven by a macroscopic forcing that is uniform at the scale of the inter-swimmer distance. Experimentally, these active colloids have been observed to reach speeds ranging from a few to tens of diameters per second \cite{Volpe11,buttinoni13,Howse07}. Promising experimental results and theoretical predictions have motivated research into possible biomedical and engineering applications such as cargo transport for targeted drug delivery \cite{Wang13b,Popescu11} and micromachines \cite{Catch05}.

These colloids are fundamentally out of equilibrium as they continuously convert physico-chemical energy of their environment into mechanical work. Hydrodynamic and chemical interactions of such self-propelled particles may lead to complex collective dynamics including clustering \cite{Theurkauff12,Palacci13,Wang13} or richer patterns~ \cite{Zottl,Illien17,saha14,cohen14}, which can often be linked to instabilities of homogeneous suspensions~\cite{Golestanian12,liebchen17}.  Individually non-motile particles can further achieve self-propulsion by forming stable chemically-inhomogeneous clusters~ \citep{soto14,soto2015}, a phenomenon which was recently also observed in experiments~\cite{wykes16,schmidt2018}.

To achieve self-propulsion, autophoretic colloids must set the surrounding fluid into motion which fundamentally requires an asymmetric distribution in the solute concentration at their surface. So far, three different mechanisms have been identified to achieve such symmetry-breaking of the concentration field: (i) an asymmetric chemical patterning of the surface (e.g. Janus particles, \cite{Howse07,Golestanian07}) (ii) an asymmetric shape of the chemically-homogeneous colloid~\cite{shklyaev2014,michelin2015a} and (iii) an instability resulting from the non-linear advective coupling between the solute dynamics and the flow motion~\cite{Michelin13,Izri14}. The former two are fundamentally associated with the particle design, and are built into its architecture. The latter, in contrast, arises spontaneously from the destabilization of a non-motile isotropic state. 

\begin{figure}[t]
\begin{center}

\includegraphics[height=2.5 cm]{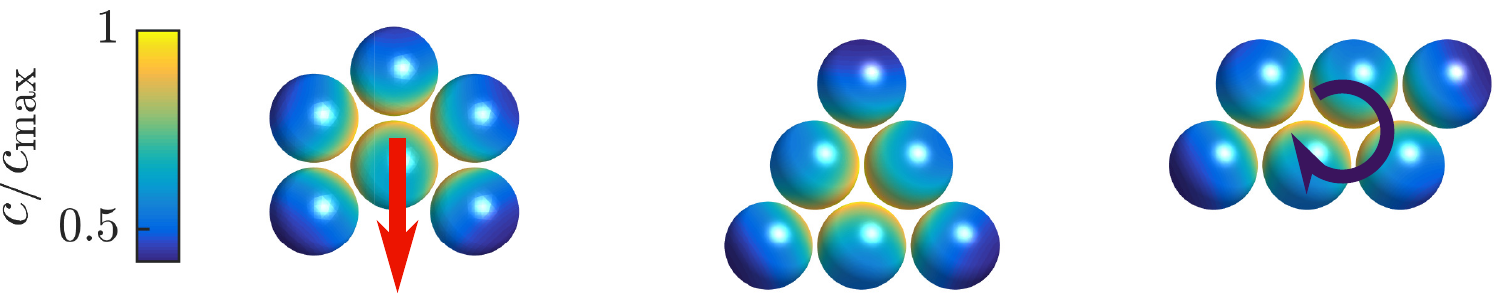}
\caption {Stable planar arrangements of $N=6$ attracting phoretic particles. Colorbar shows the relative surface concentration. The asymmetry of the geometry sets whether the cluster translates (left) remains stationary (center) or rotates (right). }
\label{fig:cluster6}
\end{center}
\end{figure}

The purpose of the present work is to introduce and characterize a fourth route to symmetry-breaking of the concentration field and self-propulsion, wherein geometrically- and chemically-isotropic particles interacting with diffusive solutes do not propel on their own but instead gain locomotion from forming asymmetric clusters. All particles are identical here, which is a fundamentally different situation from the assembly of chemically-inhomogeneous molecules from a mixture of two different types of particles~\citep{soto14,soto2015}. Identical isotropic phoretic particles which can attract (for $\mathcal{A}$ and $\mathcal{M}$ of opposite signs) each other because they generate radial concentration gradients which induce a phoretic drift on the other particles. This phoretic attraction combined with steric constraints enable only a discrete set of stable configurations that may display a geometric asymmetry (Figure~\ref{fig:cluster6}), which is a sufficient ingredient for self-propulsion of this assembly~\cite{michelin2015a}.

This collective self-propulsion is therefore intimately linked to the exact geometry of the particle assembly, which itself results from the dynamic phoretic clustering of the particles. An essential goal of the present work is therefore to characterize the statistical properties of the particle arrangement arising from the clustering process, and therefore requires a careful modelling of this dynamics.

Note that self-propulsion and collective dynamics of chemically-isotropic and individually non-motile particles was also observed for colloidal particles trapped at a fluid-fluid interface~\cite{Dom16}; then fluid motion resulted from the Marangoni stresses at the free surface rather than a direct hydrodynamic forcing by each particle as considered here.

The general problem for $N$ particles is first presented in Section~\ref{sec:general}. In Section~\ref{sec:2p}, the collective dynamics of two identical isotropic particles is considered in detail and their relative motion is computed analytically. Such analytic solutions are not available for larger number of particles and Section~\ref{sec:model} focuses on the modeling of the clustering dynamics and cluster self-propulsion. A reduced-order model, validated with full numerical solution using a regularized Boundary Element Method (BEM), is then used in Section~\ref{sec:results} to determine the statistics of formation of different clusters, their velocity and the resulting mean properties and their evolution with $N$. Finally, Section~\ref{sec:noise} analyses the effect of noise and Brownian motion on these results and conclusions are finally drawn in Section~\ref{sec:conclusions}.

\section{Collective dynamics of $N$ isotropic phoretic particles}
\label{sec:general}
The dynamics of $N$ identical spherical particles is considered in an unbounded fluid of viscosity $\eta$ and density $\rho$. The surface of the particles interacts with a chemical solute suspended in the fluid phase; following the classical continuum framework~\cite{anderson1989,Golestanian05,Golestanian07,michelin2014}, this local interaction results in an effective slip velocity $\mathcal{M}(\mathbf{I}-\nb\nb)\cdot\grad C$ where $C(\xb)$ is the surface solute concentration, $\nb$ the local normal unit vector to the surface and $\mathcal{M}$ the phoretic \emph{mobility}, which is a signed characteristic of the solute-particle interaction. The second physico-chemical property of the particles' surface is its \emph{activity}, namely its ability to alter the solute concentration (e.g. through catalytic reactions). For simplicity, this activity is modeled here as a fixed-flux emission ($\mathcal{A}>0$) or absorption ($\mathcal{A}<0$) of solute, but this framework could easily be extended to more complex chemical kinetics~\cite{ebbens2012,michelin2014,sharifimood2013,ibrahim2017}. The solute diffuses in the fluid domain with diffusivity $\kappa$, and its far-field concentration is noted $C_\infty$. The particles considered here are geometrically and chemically identical and homogeneous, i.e. $\mathcal{A}$ and $\mathcal{M}$ are constant and uniform for all spherical particles, which are also of the same radius.

The radius $a$ of the particles is chosen as the characteristic length scale. The relative solute concentration is then defined as $c=(C-C_\infty)/(|\mathcal{A}|a/\kappa)$, and characteristic velocity and pressure are chosen as $|\mathcal{A}\mathcal{M}|/\kappa$ and $|\mathcal{A}\mathcal{M}|\eta/\kappa a$, respectively. In the following, all quantities are non-dimensionalized using these characteristic scales. 

We assume that the particles are small enough so that inertial and advective effects are negligible on the fluid and solute transports (i.e. the Reynolds number, $\Rey=\rho |\mathcal{A}\mathcal{M}|a/(\eta \kappa)$ and P\'eclet numbers, $\Pe=|\mathcal{A}\mathcal{M}|a/\kappa^2$ are both small). Thus, in this purely diffusive limit, the solute concentration field obeys the steady state diffusion equation,    
\begin{equation}
\nabla^2 c=0
\label{laplace}
\end{equation}
with boundary conditions
\begin{equation}
c(r \to \infty) = 0,\qquad 
\left.\mathbf{n} \cdot  \nabla c\right|_{\mathcal{S}_j} = -A\label{eq:laplacebc}
\end{equation}
where $A=\mathcal{A}/|\mathcal{A}|=\pm 1$ is the dimensionless surface activity on the surface ${\mathcal{S}_j}$ of each particle $j$. Since advection of the solute by the fluid is neglected here, the solute dynamics can be solved for independently and the fluid dynamics problem can then be obtained in a second step using the dimensionless Stokes flow equations.
\begin{equation}
\nabla^2 \mathbf{u}=\nabla p,\;\;\ \text{and} \;\;\; \nabla \cdot \mathbf{u} = 0
\label{Stokes}
\end{equation} 
with boundary conditions
\begin{align}
\left.\ub\right|_{\mathcal{S}_j} & =\Ub_j+\Omegab_j\times\Rb_j+\tilde\ub_j \\
\tilde{\mathbf{u}}_j & =M(\mathbf{I}-\mathbf{n} \mathbf{n})\cdot \nabla c \qquad \textrm{   and   } \quad \mathbf{u} (r \to \infty)  \to 0\label{eq:mobility}
\end{align}
where, $M=\mathcal{M}/|\mathcal{M}|=\pm 1$ is the dimensionless mobility, $\Rb_j$ is the position of the center of particle $j$, and $\mathbf{U}_j=\dot{\Rb}_j$ and $\Omegab_j$ are its translation and rotation velocities. The latter quantities are determined uniquely by imposing that each particle $j$ remains force- and torque-free, i.e. for each $j$
\begin{align}
&\int_{\mathcal{S}_j}\sigmab\cdot\nb\,\dd S+\Fb_j=0,\\
&\int_{\mathcal{S}_j}(\rb-\Rb_j)\times(\sigmab\cdot\nb)\,\dd S+\mathbf{T}_j=0,
\end{align}
with $\sigmab = -p \mathbf{I} + (\nabla \mathbf{u} + \nabla {\mathbf{u}}^T) $, the stress tensor. $\Fb_j$ and $\mathbf{T}_j$ are the contact (e.g. steric) forces applied on particle $j$ in a cluster. When the particles are independent, they are force and torque-free, i.e. $\Fb_j=0$, $\mathbf{T}_j=0$ for each value of $j$. When the particles form a cluster, their overlap is prevented by steric and other short-ranged effects. Thus, in this case, the particles are individually not force- and torque-free, but rigidly-bound. However, the system of equations can be closed by imposing that the system as a whole is force- and torque-free ($
\displaystyle\sum_j\mathbf{F}_j =0, \;\; \sum_j \mathbf{T}_j = 0$).
In the following, we assume that steric interactions between particles generate only central forces so that $\mathbf{T}_j=0$.

Indeed, for spherical and isotropic particles with negligible solute advection, a rotation of any particle around its center leaves the solute concentration unchanged. In the following, we therefore solely focus on the translation velocities of the particles, $\mathbf{U}_j$ (note that $\Omegab_j$ may not be zero, but is set to satisfy the condition that each particle remains torque-free). 

\begin{figure}[t]
\begin{center}
\begin{tabular}{c c}
\includegraphics[height=4.5cm]{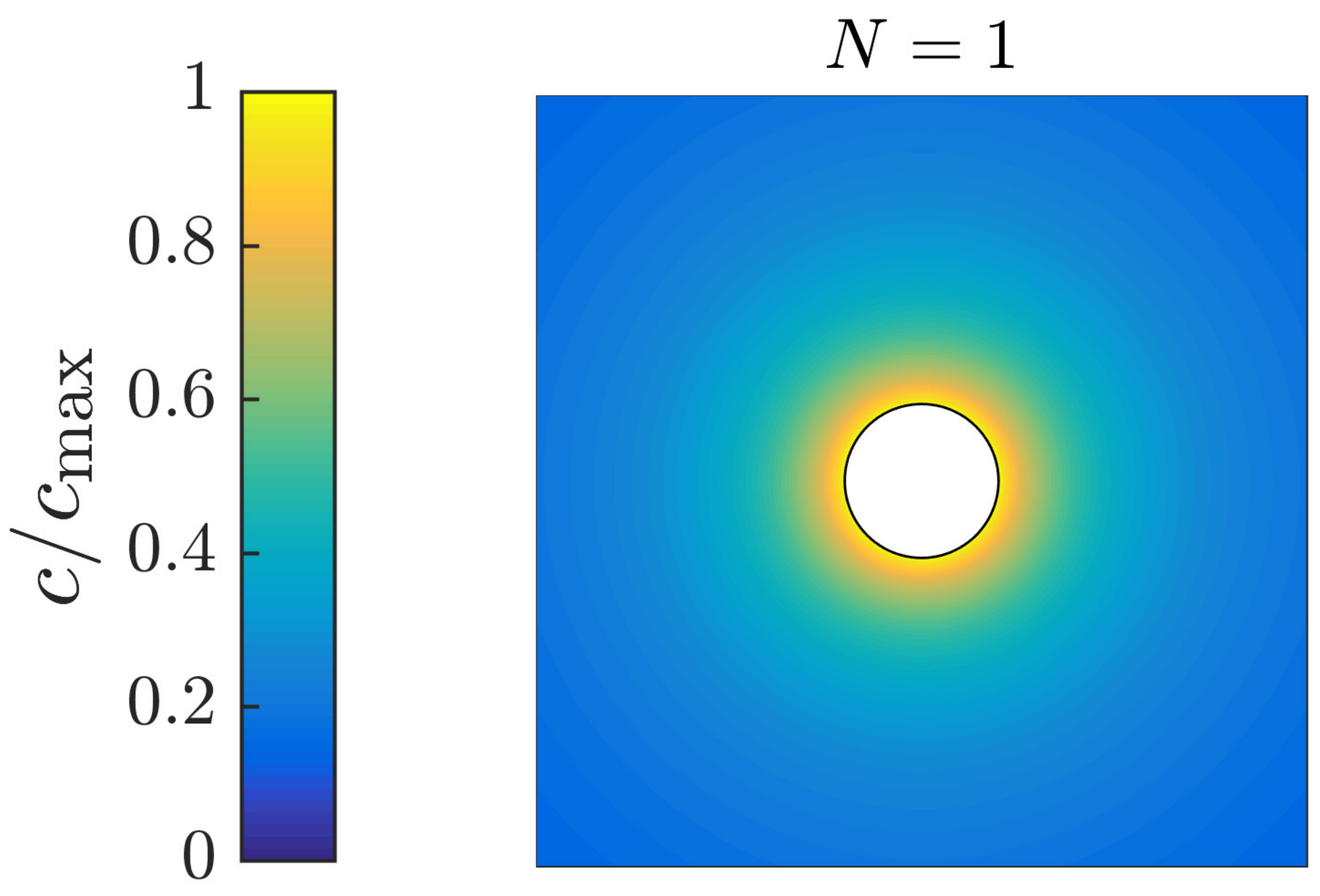}
\includegraphics[height=4.52cm]{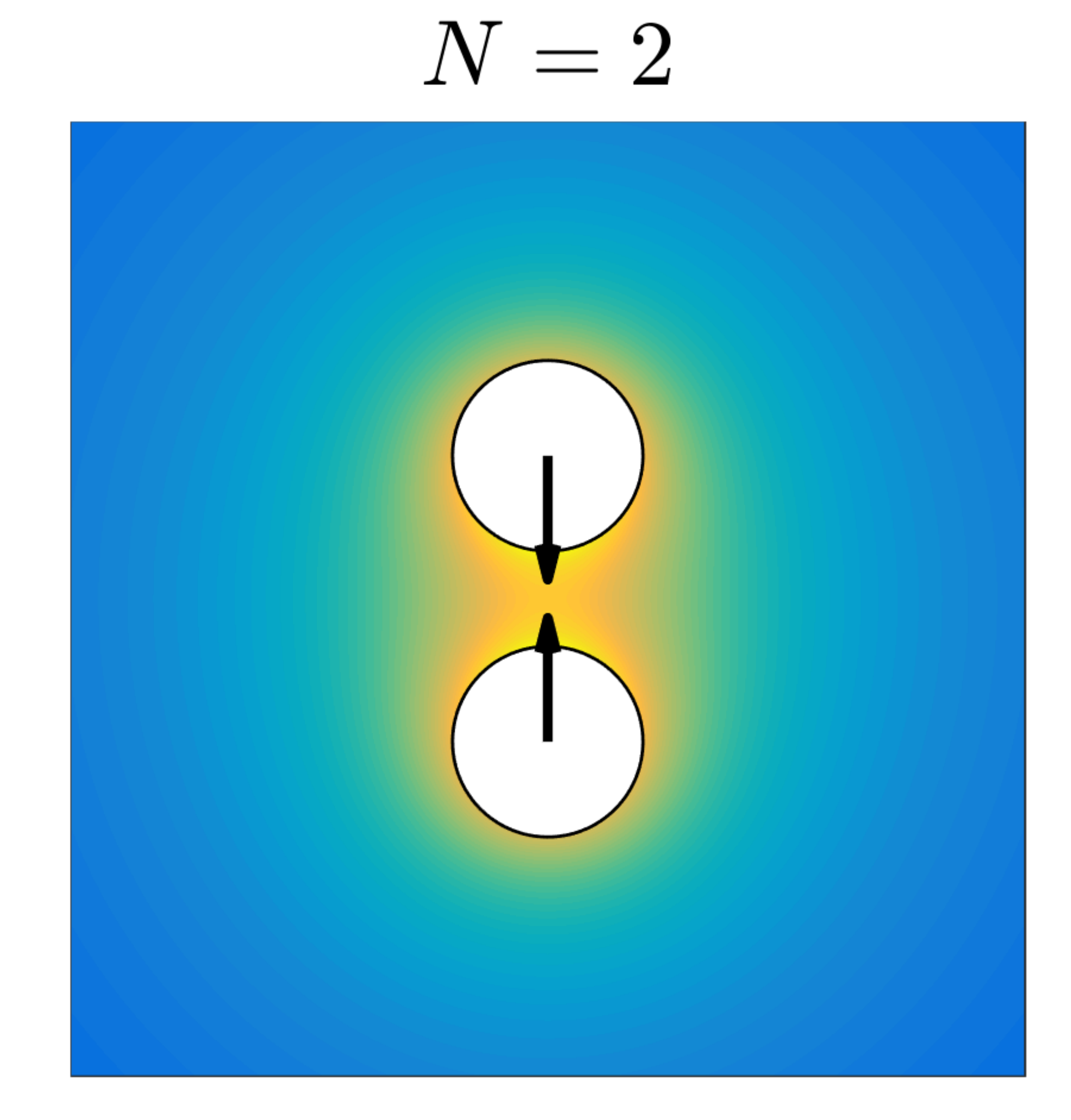}
\end{tabular}
\caption{Relative concentration field around (a) a single isotropic phoretic particle and (b) two identical isotropic phoretic particles ($A=1$ and $M=-1$). In (a), the isotropic concentration field leads to no fluid motion nor propulsion. In (b), the asymmetric concentration field around each particle leads to a mutual attraction.}
\label{fig:isolated}
\end{center}
\end{figure}

For a single isolated particle ($N=1$), the concentration field is purely isotropic $c(r)=A/r$ and does not generate any surface gradient (or slip velocity) along its surface; a single isotropic phoretic particle is therefore unable to self-propel despite its chemical activity (see Figure~\ref{fig:isolated}a).

However, the symmetry of the concentration field is broken by the presence of another identical particle in the system (Figure~\ref{fig:isolated}b). The surface concentration gradients lead to a mutual phoretic attraction (or repulsion, depending on the relative sign of their surfaces' properties, $A$ and $M$) between the particles along their axis of symmetry, $\mathbf{e_z}$. The resulting  phoretic interactions can be computed and characterized analytically for two particles, and this is the focus of the next section.

\section{Phoretic clustering of two identical particles}
\label{sec:2p}
\subsection{Intuitive model}
To understand symmetry-breaking and the resulting clustering, we first take a look at the far-field limit when the distance between the particles $d$ is much greater than their radius ($a/d \ll 1$). In this limit, in addition to its own chemical field, particle $i$ is exposed at leading order to the source field $A/r_j$ of particle $j\neq i$, which creates a locally uniform concentration gradient
\begin{equation}
\left.\grad c\right|_{\Rb_i}=-\frac{A(\Rb_j-\Rb_i)}{d^3}
\end{equation}
with $d=|\Rb_2-\Rb_1|$. In response to this gradient, particles experience equal and opposite phoretic drift velocities along their axis of symmetry~\cite{anderson1989}
\begin{equation}
\mathbf{U}_{1} = \frac{AM a^2}{d^2}\; \mathbf{e_z}, \;\;\;\; \mathbf{U}_{2} = -\frac{AM a^2}{d^2}\; \mathbf{e_z}
\label{far}
\end{equation}
with $\eb_z=(\Rb_1-\Rb_2)/d$. Phoretic attraction (resp. repulsion) therefore arises for $AM=-1$ (resp. $AM=1$). We shall henceforth focus exclusively on the attractive case ($AM=-1$) which will be responsible for clustering. When the far-field assumption does not hold, this velocity must be corrected to account for confinement effects in both the diffusion and hydrodynamic problems. It is yet amenable to exact analytical calculation using classical results for Laplace's and Stokes' equations in bispherical coordinates~\cite{stimson1926,michelin2015a,reigh2015}.

\subsection{Analytical solution for arbitrary separation}
The clustering dynamics of two particles is solved analytically for arbitrary distance $d$ using bispherical coordinates $(\tau,\eta,\phi)$ defined from the cylindrical coordinate system $(\rho,\phi,z)$ as 
\begin{equation}
\rho = \frac{\alpha \sqrt{1-\mu^2}}{\cosh \tau - \mu}, \;\;\; z=\frac{\alpha \sinh \tau}{\cosh \tau - \mu}, \qquad \text{with } \mu = \cos \eta
\end{equation}
A constant value of $\tau$ corresponds to a sphere of radius $\alpha/\sinh|\tau|$ with center located at $\alpha/\tanh\tau$. Hence, the unit vectors, $\mathbf{e_\tau}$ and $\mathbf{e_\mu}$ are respectively, normal and tangential to a spherical surface. Here both spheres have the same unit radius, and correspond to $\tau=\pm \tau_0$. The positive constants $\alpha$ and $\tau_0$ are  determined so that
\begin{equation}
d=2\alpha \cosh\tau_0,\qquad \alpha=\sinh\tau_0. 
\end{equation}
The general solution to Laplace's equation for concentration $c$, that decays in the far-field, is obtained within this framework~\cite{stimson1926} as
\begin{align}
c = \sqrt{\cosh\tau-\mu}\sum_{n=0}^\infty & c_n(\tau) L_n(\mu), \quad c_n(\tau)=C_n \cosh\left[\left(n+\frac{1}{2}\right)\tau\right]
\end{align}
where $L_n(\mu)$ is the Legendre polynomial of degree $n$. Enforcing the surface flux boundary condition, Eq.~\eqref{eq:laplacebc}, 
\begin{align}
-\frac{n}{2n-1}c_{n-1}'(\tau_0) & +c_n'(\tau_0)\cosh\tau_0 -\frac{n+1}{2n+3}c_{n+1}'(\tau_0)  +\frac{\sinh\tau_0}{2}c_n(\tau_0)=\alpha A\sqrt{2}\ee^{-(n+1/2)\tau_0}
\end{align}
determines the coefficients $C_n$ uniquely \citep{michelin2015a}.
The surface slip velocity is then determined from the concentration distribution on the surface \cite{michelin2015a}, 
and the reciprocal theorem for Stokes flow is used to compute the axial velocities $\mp U$ of particles $1$ and $2$~\citep{stone1996,michelin2015a}:
\begin{equation}
U=\frac{1}{2F^*}\int_{\mathcal{S}_1,\mathcal{S}_2}\tilde\ub\cdot\sigmab^*\cdot\nb\,\dd S,\label{eq:vel_recipr}
\end{equation}
where $\sigmab^*$ is the hydrodynamic stress tensor of the auxiliary flow field corresponding to the solution of Stokes equations around \emph{rigid} spheres $\mathcal{S}_1$ and $\mathcal{S}_2$ translating with velocities $\pm U^*\eb_z$, in response to an outer force $\mp F^*\eb_z$. This auxiliary flow field can be expressed as 
\begin{equation}
\ub^* =-\frac{(\cosh\tau-\mu)^2}{\alpha^2}\pard{\psi^*}{\mu}\eb_\tau+\frac{(\cosh\tau-\mu)^2}{\alpha^2\sqrt{1-\mu^2}}\pard{\psi^*}{\tau}\eb_\mu
\end{equation}
with the streamfunction $\psi^*$ given by~\cite{stimson1926,michelin2015a}
\begin{align}
\frac{\psi^*(\tau,\mu)}{U^*} &= (\cosh \tau - \mu)^{-3/2} \sum_{n=1}^\infty (1-\mu^2) L'_n(\mu) U_n(\tau),\\
U_n(\tau) &= \; \beta_n \sinh\left[\left(n+\frac{3}{2}\right)\tau\right] + \gamma_n \sinh\left[\left(n-\frac{1}{2}\right)\tau\right].\label{eq:un}
\end{align}
The no-slip boundary conditions on the spheres' surface write as
\begin{align}
\psi^*(\tau_0,\mu)&=-\frac{\alpha^2 U^*(1-\mu^2)}{(\cosh\tau_0-\mu)^2},\\
 \frac{\partial \psi^*}{\partial \tau} (\tau_0,\mu) & =  \frac{\alpha^2 U^*(1-\mu^2)\sinh \tau_0}{(\cosh \tau_0-\mu)^3},
\end{align}
and appropriate projections on the Legendre polynomials provide an explicit determination for $\beta_n$ and $\gamma_n$. The shear stress distribution on the surface is then determined as~\cite{michelin2015a}
\begin{align}
\sigma^*_{\tau\mu}(\tau,\mu) = &\frac{1}{\alpha^3\sqrt{1-\mu^2}} \pard{}{\tau}\left[(\cosh\tau-\mu)^3\pard{\psi^*}{\tau}\right]  -\frac{\sqrt{1-\mu^2}}{\alpha^3}\pard{}{\mu}\left[(\cosh\tau-\mu)^3\pard{\psi^*}{\mu}\right],
\label{eq:shstress}
\end{align}
and the hydrodynamic force on each sphere is obtained directly as~\cite{stimson1926}
\begin{align}
F^* = \frac{2\pi\sqrt{2}}{\alpha}\sum_{n=1}^\infty n(n+1)(\beta_n+\gamma_n).
\label{eq:force}
\end{align}

\begin{figure}[t]
\begin{center}
\includegraphics[width=.7\textwidth]{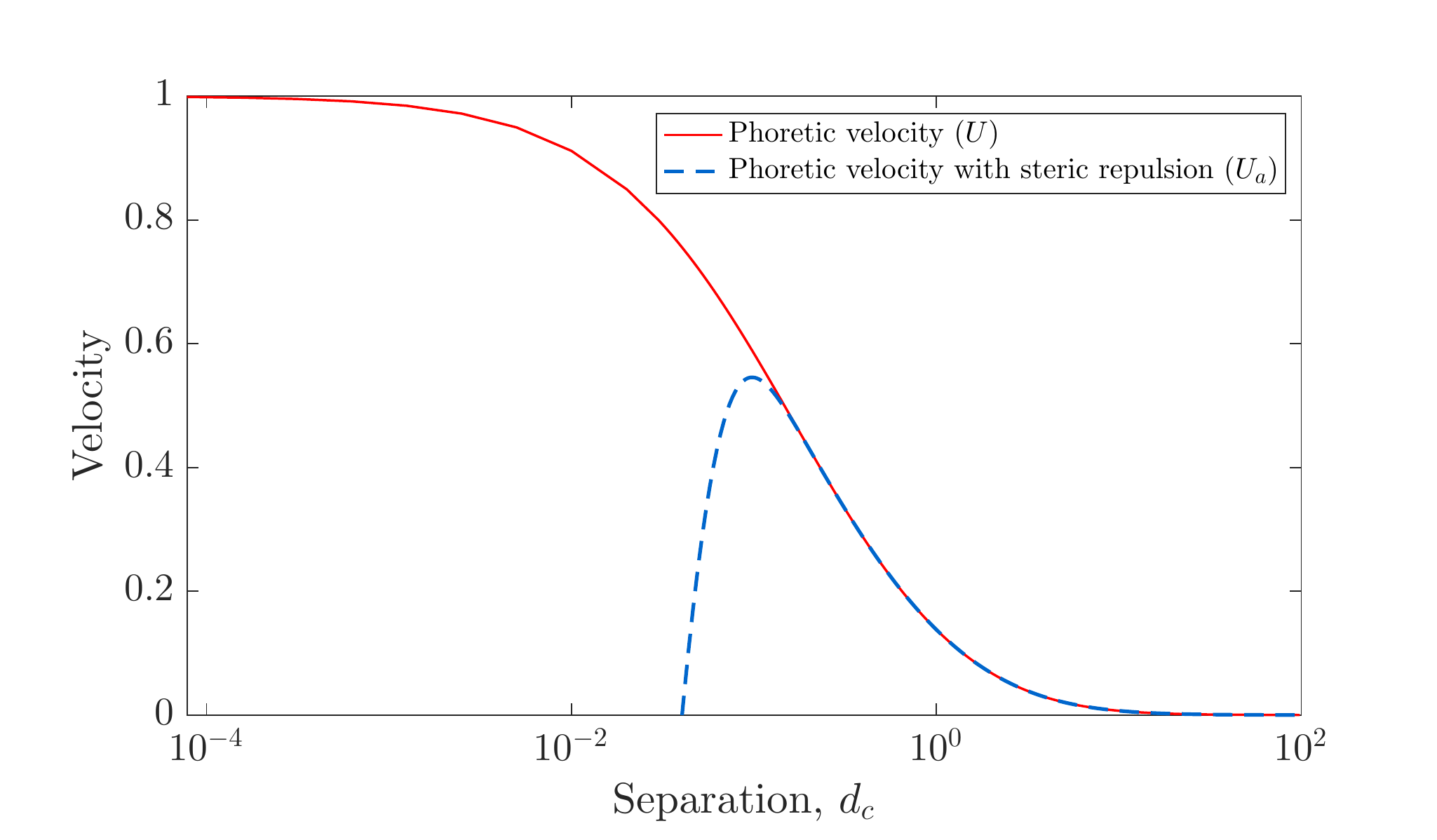}
\caption {Clustering velocity of a catalytic particle of unit radius. The solution converges to $1/d^2$ decay, as expected in the far-field ($d_c \to \infty$); In the case with repulsion velocity, $C=35$, $\delta=25$ and $d^{*}=1.95$ were chosen in Eq.~\eqref{eq:complete}.}
\label{velfn}
\end{center}
\end{figure}

Note that, due to action-reaction symmetry, this mutual attraction/repulsion velocity depends only on the distance $d$ between the particles. Its value is hence computed numerically at various separation, $d_c=d-2 a$  and is plotted in Figure~\ref{velfn}.

The phoretic slip model~\cite{anderson1989}, Eq.\eqref{eq:mobility}, at the heart of the present modeling likely breaks down when $d_c\sim \lambda$ with $\lambda$ the thickness of the interaction layer where solute-particle interactions are significant. Yet, the phoretic slip model remains valid everywhere except within a small lubrication region between the particles which is only expected to introduce a subdominant correction to the present predictions.

The clustering velocity converges to a unit value when the particles are in contact.  Hydrodynamic (lubrication) and phoretic effects are thus insufficient to avoid particle contact in finite time~\cite{yariv2016}; short-ranged interaction forces must therefore be added to account for intermolecular repulsive forces between the interacting surfaces preventing particle overlap (i.e. steric effect). To preserve the previous velocity formulation, this repulsive effect is therefore included as an additional repulsive velocity to mimic this behaviour, in the form of a smoothed step-function. The complete expression for the clustering velocity is hence given by
\begin{equation}
U_a(d)= U(d) -C\;(1-\tanh(\delta\;(d-d^{*}))),
\label{eq:complete}
\end{equation}
where $C$, $\delta$ and $d^{*}$ are chosen appropriately so that the particles have negligible separation after clustering ($d_c \ll a$). This clustering velocity is a function solely of the particles' distance, shown in Figure~\ref{velfn}, and the dynamics of the two particles can therefore be described by an over-damped deterministic Langevin dynamics within an interaction potential:
\begin{equation}
\totd{\Rb_j}{t} =-\pard{\mathcal{E}}{\Rb_j},\quad \textrm{with    }\mathcal{E}(\Rb_1,\Rb_2)=\mathcal{E}_{2p}(|\Rb_1-\Rb_2|),
\label{langevin}
\end{equation}
and 
\begin{equation}
\pard{\mathcal{E}_{2p}}{d} =-U_a(d).
\label{langevin2}
\end{equation}
This potential $\mathcal{E}$ is a measure of the stability of the two-particle cluster, and the system evolves at each instant down the direction of its steepest gradient.

For two particles, hydrodynamic and phoretic interactions introduce an asymmetry in the system that enables the motion of individual particles. Yet, the center of mass of the arrangement remains stationary for identical particles, since the final cluster (i.e. dimer) is front-back symmetric. For identical particles, the emergence of global self-propulsion is therefore restricted to $N>2$.

\section{Modeling the clustering and collective motion of $N$ particles}
\label{sec:model}

While an analytical solution was amenable for $N=2$ particles, this is not the case for larger $N$. Analytical methods, based for example on the Method of Reflections (MoR), can still be used to provide asymptotic approximations in particular for large separation distances between the particles (see Appendix~\ref{app:mor}). Such approximations provide valuable qualitative and quantitative insight, but they fail to provide a description of the near-field interactions. These could nevertheless be studied using asymptotics through lubrication analysis~\cite{yariv2016}. Yet, in order to obtain a complete description of both the chemical and hydrodynamic problems for arbitrary arrangements, it is necessary to resort to numerical simulations of the Laplace and Stokes' problems. In problems involving fluid-solid interactions, commonly employed numerical techniques include finite element and immersed boundary methods \cite{Randall} and its variations \cite{Wagner,Liu}. 

Boundary element methods (BEM), based on classical boundary integral representations of the flow field, have also found an important niche in solving Stokes flow problems~\cite{Pozrikidis}. However, their implementation for phoretic problems has only been recent ~\cite{uspal14,montenegrojohnson2015}. Boundary Element Methods solve the Laplace~\eqref{laplace} and Stokes~\eqref{Stokes} equations outside a set of rigid particles by using the fundamental integral representation of the solutions to these harmonic and bi-harmonic equations, in terms of their values and normal gradients on the bounding surfaces only. Such methods are therefore particularly well-suited for phoretic problems in Stokes flows since the coupling of the concentration and velocity fields occurs only on the particles' surfaces. 

Because they represent the solution as the superposition of fundamental singularities (e.g. sources and Stokeslet), classical Boundary Element Methods for Laplace and Stokes problems involve singular kernels which require a separate analytical treatment of the singular contributions and precise quadrature techniques. An alternative is to regularize the singularities, for example using the method of regularized stokeslets for Stokes flows ~\cite{Cortez2001, Cortez2005, David2009}. 
 
\subsection{Regularised Boundary Element Method}
\label{subsec:BEM}

The concentration as well as the flow fields, being harmonic functions, can be evaluated at any point in the domain using boundary integral representations. BEMs are numerical approximations to these surface integrals obtained by discretising the surface into a collection of \textit{elements}. The surface concentration field is evaluated by distributing regularized sinks and source dipoles on the surface of the particle while the surface flow velocities are computed by distributing regularized stokeslets and stresslets. These regularised singularities are collocated at discrete points on the surface (called \textit{nodes}) which are element vertices.

We follow the computational framework of regularised BEM developed for phoretic problems by Montenegro-Johnson \emph{et al.}~\cite{montenegrojohnson2015} to generate mesh and quadrature routines for evaluation of surface integrals. The surface of each phoretic particle is discretised into 3072 piecewise-quadratic triangular elements (1538 linear nodes). Surface unknowns are discretized as linear functions over each element for accurate computation of the concentration field and traction on the surface. The method begins with the regularized boundary integral equation for the concentration field in response to the flux forcing on the active particles~\cite{montenegrojohnson2015}
\begin{align}
\int_{V_f}c(\xb)\phi^\epsilon(\xb,\xb_0)\mathrm{d}\mathcal{S}_\mathbf{x} = \int_\mathcal{S} \Big[& c(\mathbf{x})\;\mathbf{K}^\epsilon (\mathbf{x},\mathbf{x}_0) \cdot \mathbf{n}(\mathbf{x}) - \frac{\partial c(\mathbf{x})}{\partial n}\;G^\epsilon(\mathbf{x},\mathbf{x}_0) \Big]\mathrm{d}\mathcal{S}_\mathbf{x}\label{eq:regbem}
\end{align}
where $\mathbf{x}$ is any point on the surface $\mathcal S$ of the spherical particles with local normal $\mathbf{n}(\mathbf{x})$ and $\mathbf{x}_0$ is the point where concentration is to be calculated. The regularised blob function and associated kernels are given by 
\begin{equation}
\phi^\epsilon(\xb,\xb_0)=\frac{15\epsilon^4}{8\pi r_\epsilon^7},\qquad
G^\epsilon(\mathbf{x},\mathbf{x}_0) = -\frac{(2r^2+3\epsilon^2)}{8 \pi r_\epsilon^3}, \qquad K_j^\epsilon(\mathbf{x},\mathbf{x}_0) = r_j\frac{2r^2+5\epsilon^2}{8\pi r_\epsilon^5},\label{green}
\end{equation}
with $r=|\mathbf{x}-\mathbf{x}_0|$ and $r_\epsilon^2 = r^2 + \epsilon^2$.  $G^\epsilon$ and $K_j^\epsilon$ represent a regularised sink and source dipole respectively. The essence of the regularized boundary method is to express simply the left-hand side of Eq.~\eqref{eq:regbem} in terms of the local concentration $c(\xb_0)$ so as to obtain an integral equation to solve for $c(\xb_0)$ on the boundary of the particles. When $\xb_0$ is located on the surface of a particle, the left-hand side of Eq.~\eqref{eq:regbem} can be expressed as~(see Appendix~\ref{app:reg})
\begin{equation}
\int_{V_f}c(\xb)\phi^\epsilon(\xb,\xb_0)\mathrm{d}\mathcal{S}_\mathbf{x}=c(\xb_0)\left(\frac{1}{2}+\frac{\epsilon\kappa}{4}\right)+\frac{\epsilon}{4}\pard{c}{n}(\xb_0)+O(\epsilon^2)\label{eq:regeffect}
\end{equation}
with $\kappa$ the mean local curvature of the particle surface (here $\kappa=1$ for spherical particles). It should be noted that the above result is true for surfaces of arbitrary shape and that the effect of the regularization introduces two different $O(\epsilon)$ corrections to the classical $c(\xb_0)/2$ result obtained for singular boundary integral methods. These corrections are respectively proportional to the concentration and normal flux, which can be easily implemented in classical boundary integral methods frameworks \footnote{the latter was erroneously omitted in an earlier presentation of the method~\cite{montenegrojohnson2015}}. A similar correction should be implemented when $\xb_0$ is not exactly on the surface. It can however be demonstrated that these corrections scale at leading order as $O(\epsilon^4/d^4)$ and $O(\epsilon^3/d^2)$ (here $d$ is the distance to the integration surface) and are therefore negligible except in the immediate vicinity of the surface.

Once the surface concentration field has been obtained, the surface slip velocity, $\tilde{\mathbf{u}}$, is computed from the particle's phoretic mobility property. The boundary integral formulation of the Stokes flow problem and force- and torque-free condition on each particle is then solved for the flow traction and translation and rotation velocities $\Ub_j$ and $\Omegab_j$ of the different particles using \cite{Cortez2001,montenegrojohnson2015}
\begin{align}
\int_{V_f}u_j(\mathbf{x})\phi^\epsilon(\xb,\xb_0)\mathrm{d}\mathcal{S}_\mathbf{x} = &\frac{1}{8\pi}\int_\mathcal{S} \Big[S_{ij}^\epsilon(\mathbf{x},\mathbf{x}_0)\; f_i(\mathbf{x}) -u_i(\mathbf{x})\;T_{ijk}^\epsilon(\mathbf{x},\mathbf{x}_0)\; n_k(\mathbf{x})\Big]\dd\mathcal{S}_\mathbf{x}\label{eq:regstokes}
\end{align}
where $\mathbf{f}$ is the surface traction and the flow velocity at the surface, $\mathbf{u}=\tilde{\mathbf{u}}+\mathbf{U}+\Omegab \times \mathbf{x}$. Here, the Green's functions $\mathbf{S}^\epsilon$ and $\mathbf{T}^\epsilon$ represent a regularised stokeslet and its associated stress tensor, respectively:
\begin{align}
S^\epsilon_{ij}(\mathbf{x},\mathbf{x}_0) &= \frac{(r^2+2\epsilon^2)\delta_{ij} + r_i r_j} {r_\epsilon^3},\qquad T^\epsilon_{ijk}(\mathbf{x},\mathbf{x}_0) = \frac{-6 r_i r_j r_k -3\epsilon^2(r_i \delta_{jk}+r_j \delta_{ik}+r_k \delta_{ij})}{r_\epsilon^5}\cdot
\end{align}
If one is only interested in the resulting motion of the particles, such a boundary integral approach is particularly well-suited as no bulk quantity needs to be computed and remeshing the domain at each time step is unnecessary. The integral in the left-hand side of Eq.~\eqref{eq:regstokes} can be treated similarly as for the Laplace problem. Only if necessary (e.g. for plotting purposes), the surface distribution of concentration and velocity may be used to compute their bulk distribution, using the fundamental integral representations. 

\begin{figure}[t]
\begin{center}
\includegraphics[width=.43\textwidth]{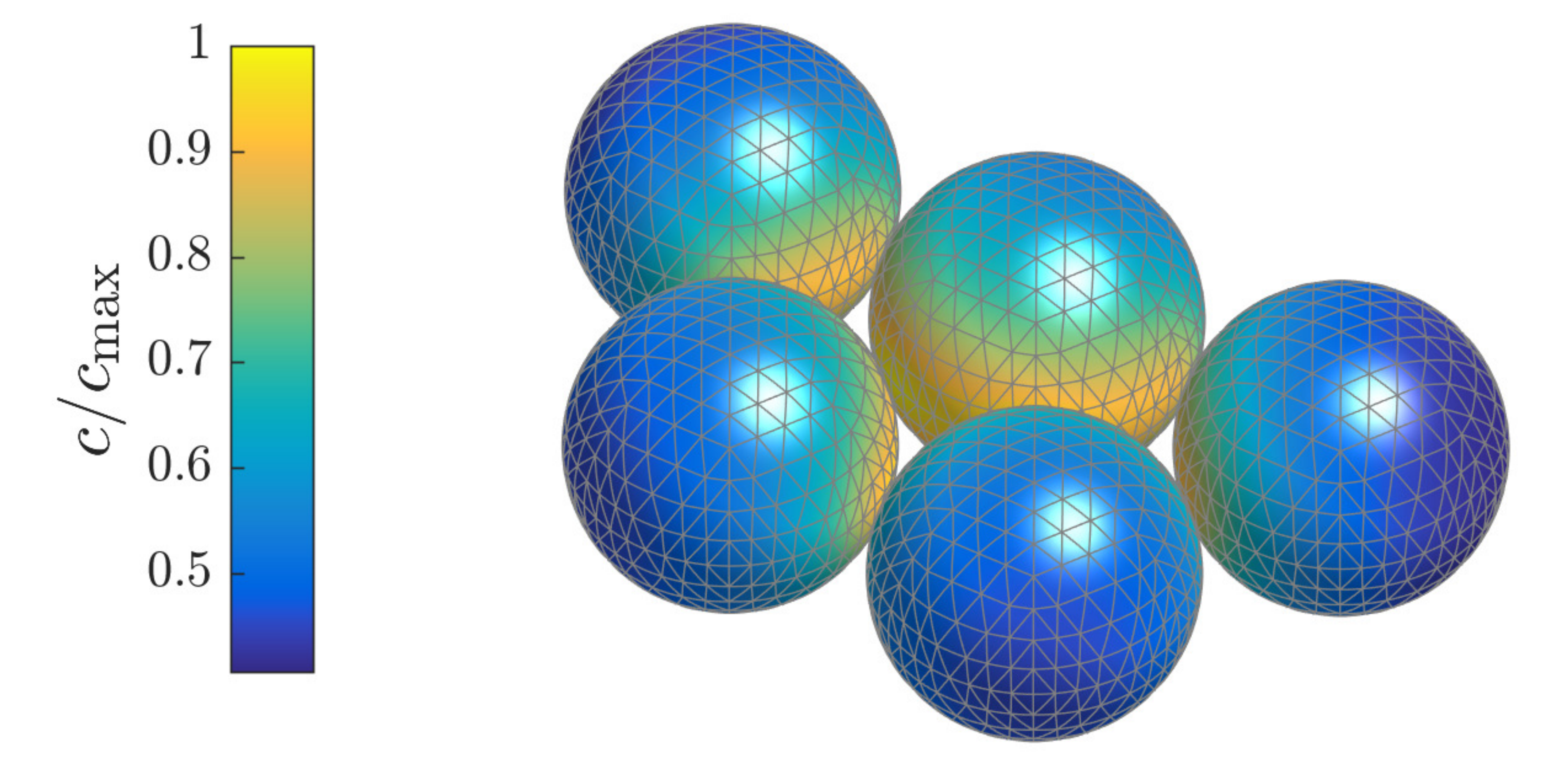}
\caption{An example of the mesh used for regularised BEM computations (coarsened to $1/4^{\text{th}}$ the total number of elements used so as to aid visualisation). The surface concentration distribution on the 5-particle cluster obtained from the simulations are also shown.}\label{fig:5pmesh}
\end{center}
\end{figure}

For simplicity, we restrict the following analysis to two-dimensional clustering and self-propulsion of the spherical particles. However, we note that the solute diffusion and hydrodynamic problem remain three-dimensional and unbounded. A variable mesh size is implemented along both polar and azimuthal directions of each sphere such that the mesh is refined within and near the plane of clustering. Further, in stable clusters, particles are arranged on a regular hexagonal lattice. Additional mesh refinement is then performed in the clustering plane near the positions of particles' contact on hexagonal lattices. This meshing provides an efficient framework to compute the swimming velocity of clustered particles (see Figure~\ref{fig:5pmesh}).

The regularization parameter $\epsilon$ must be chosen small enough to approach the true solution but large enough to avoid a singular behaviour of the integral equation; a value of $0.005$ and $0.01$ is typically used. This regularized Boundary Element Method was shown to provide accurate computations for near- and far-field dynamics of various phoretic problems~\cite{montenegrojohnson2015}. We validated here the method on the two-particle problem for which an analytical solution was obtained. Using the above mentioned meshing, an accuracy of about $1\%$ was obtained on the swimming velocity of the cluster even for particles near contact with $d_c=0.01$ and $d_c=0.02$ (about 10 times the local element size in the adapted mesh).

In the following, two different types of problems are considered and solved numerically. In the first one, the detailed kinematics of the $N$-particle system is marched in time using an adaptive two-step Adam-Bashforth method, where the time step is determined on the smallest-separation between the particles. To prevent particle overlap, a soft repulsive potential is introduced akin to that leading to the corrective relative velocity for the two-particle system. A relatively coarse and uniform mesh (512 nodes per sphere) is used to compute the velocities of the particles, for which computing the complete set of trajectories for $6$ particles initially distributed within a circle of radius $R_{\mbox{\scriptsize{max}}}=10$ requires about 12 hours using Matlab on a desktop computer. This computational cost is prohibitive for large number of simulations (see Section~\ref{subsec:pair}), hence providing a clear motivation for a reduced-order model of the clustering dynamics. These accurate BEM simulations nevertheless provide the required reference for validation of such model.

 In the second class of problems, regularized BEM are used to obtain the translation and rotation velocities of a cluster of phoretic particles (i.e. frozen relative positions). Numerically, the particle assembly is considered as rigid, and the resulting velocity of these stable clusters is computed with inter-particle separation, $d_c=0.01$ and $d_c=0.02$; a Taylor series expansion of the global swimming velocity of the cluster in terms of contact distance around $d_c=0$ is then used to extrapolate the true self-propulsion velocities of clusters when particles are in actual contact. To achieve sufficient accuracy on the global motion of the cluster (in particular resolving properly the flow field around the particles in the lubrication regions), a finer mesh is used with 1538 nodes and refinement near the regions of contact as described previously. Computation of the cluster velocity for $N=13$ yet requires $\approx 28$ GB of allocated memory, but only two computations are performed for each cluster shape.

\subsection{Clustering vs. Self-propulsion}

The numerical methods outlined in the previous section are now used to compute the dynamics of $N$ identical spherical particles of unit radius, with chemical properties resulting in a mutual phoretic attraction i.e. $AM=-1$. For simplicity, their initial arrangements (and hence dynamics) are restricted to two dimensions, the solute diffusion and fluid motion remaining fully three-dimensional. The present approach and formalism could nevertheless be applied directly to 3D motion. Particles are initially arranged randomly in a plane; under the effect of phoretic attractions, complex non-equilibrium clustering dynamics are observed leading to the formation of a stable rigid planar assembly, held together by the balance between phoretic attraction and short-ranged repulsive forces. These clusters are stable in the classical sense that any slight two-dimensional perturbation would return them back to their original configuration. Using optimal packing arguments (attracting phoretic particles tend to minimize their relative distance), particles in such stable clusters  are found to arrange on a regular hexagonal lattice where the particles have no relative degree of freedom left i.e. each particle cannot move without separating from its adjacent particle in the lattice.

\begin{figure*}[t]
\begin{center}
\begin{tabular}{c}
\includegraphics[width=.7\textwidth]{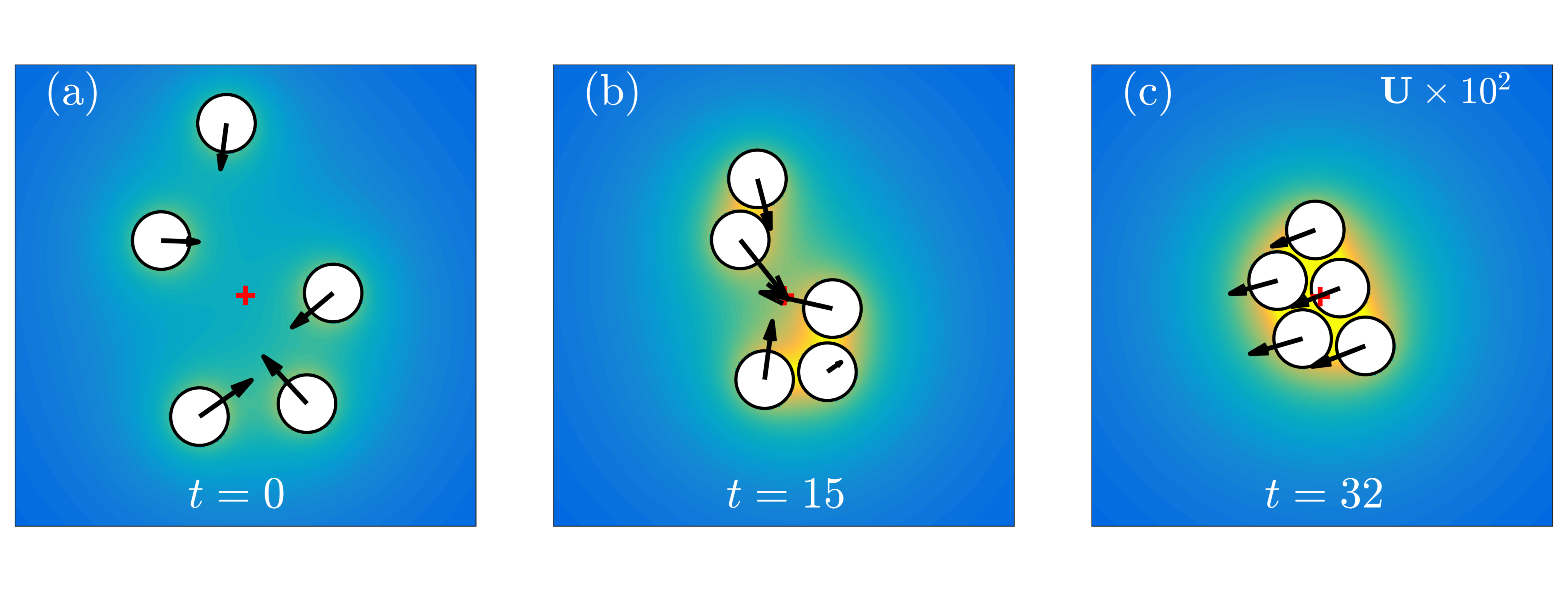}\\
\includegraphics[width=0.58\textwidth]{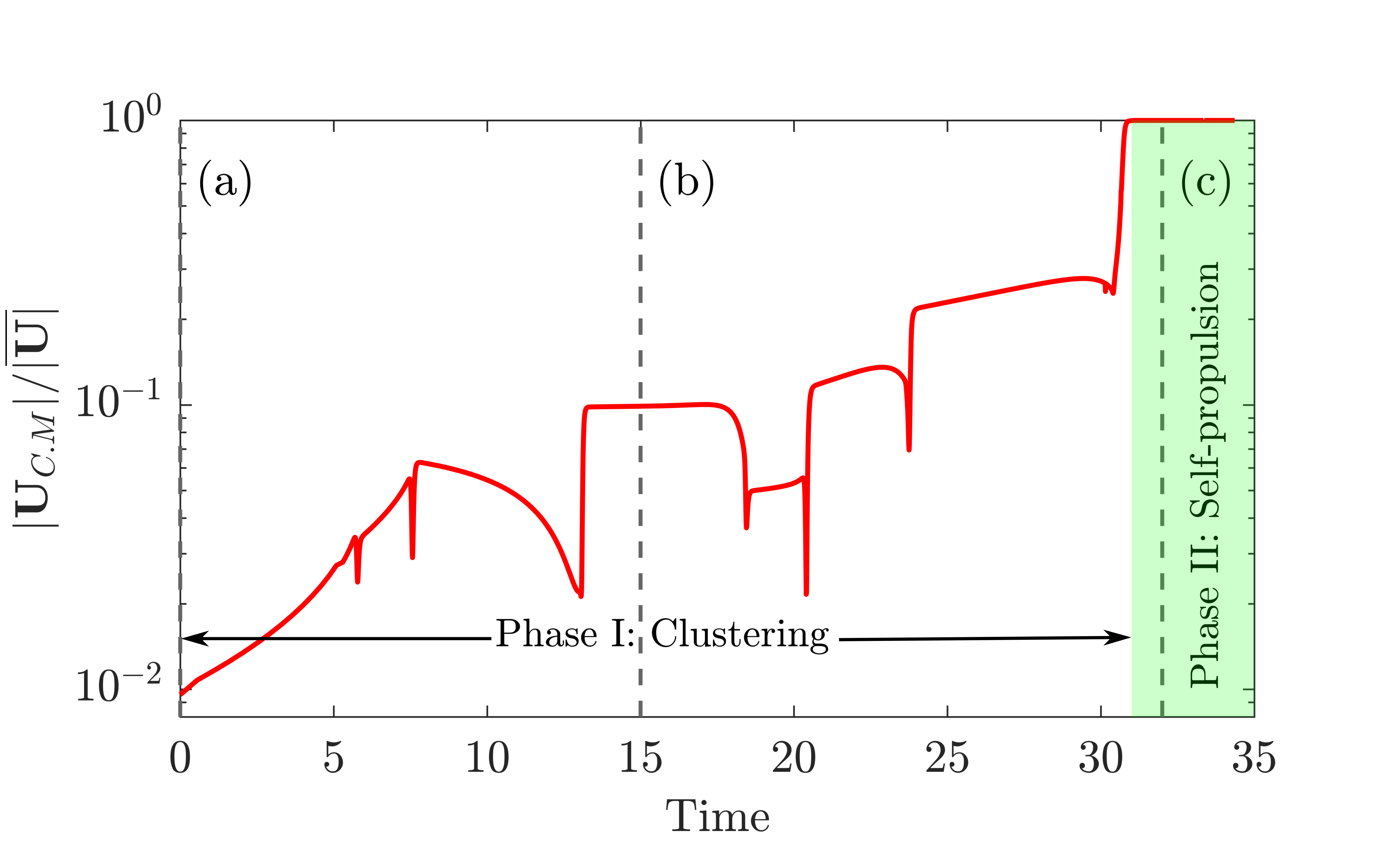}
\end{tabular}
\caption {(Top) Clustering dynamics of 5 isotropic particles ($A=1,M=-1$). The velocity of each particle is shown by black arrows and the color indicates the concentration field. The red cross indicates the position of the center of mass. (Bottom) Ratio of the magnitude of the center of mass velocity to the average individual velocity magnitude of the 5 particles. A stable cluster is formed when there is no relative motion between the center of mass and individual particles (green shade). The vertical dashed lines correspond to the three snapshots above.}
\label{fig:cluster_vs_selfprop}
\end{center}
\end{figure*}

An example of such dynamics is shown on Figure~\ref{fig:cluster_vs_selfprop}. Initially, the asymmetric concentration field generated around each particle by its neighbours drives its motion. The general direction of each particle tends to cluster them around the geometric center of the assembly (referred to in the following simply as ``center of mass'', although no inertial process is involved here). Initially and up until the particles form a rigid cluster, the velocity of the center of mass of the particles is noticeably at least one order of magnitude smaller than the velocity of individual particles (see Figure~\ref{fig:cluster_vs_selfprop}): the center of mass of the system remains effectively stationary during that phase. This is somewhat unsurprising given the opposite (attractive) velocities induced by the particles on each other as observed in the 2-sphere case.

Yet, once a stable cluster is formed and particles experience no relative motion, the global velocity is interestingly non-zero (albeit small): particles swim as a cluster. This collective self-propulsion arises from the geometric asymmetry of the formed cluster which generates an asymmetric concentration field and phoretic forcing on the assembly, a phenomenon that was recently characterized for rigid systems of homogeneous properties and various shapes \cite{michelin2015a,shklyaev2014}.

This transition clearly decomposes the dynamics of $N$ phoretic particles into two different regimes, that differ fundamentally in the relative magnitude of the mean and relative velocities of the particles. In the clustering phase (Phase I), particles are individually force-free hydrodynamically. In the self-propulsion phase (Phase II), they are rigidly-bound by the balance of phoretic attraction and repulsive internal forces that prevent their overlap, and the hydrodynamic force on each particle is now non-zero. This profoundly modifies each particle's hydrodynamic signature. The results of the numerical simulation of the full Laplace and Stokes problems show that (i) Phase I is characterized by a global velocity (i.e. center of mass) that is negligible in front of individual motion and (ii) this global velocity is significantly enhanced in Phase (II) when the particles are rigidly-bound. 

These two features are confirmed by computing the asymptotic expansion of the particles' velocity during each phase using the \textit{method of reflections}~\cite{kimkarrila} (see Appendix~\ref{app:mor}). When the particles are hydrodynamically force-free individually (i.e. not in contact, Phase I), the translation velocity of particle $j$ is obtained as
\begin{align}
\Ub_j^\textrm{free}=-AM\sum_{\substack{j,k=1\\k\neq j}}^N\frac{\eb_{jk}}{d_{jk}^2}
+\frac{5AM}{2}\sum_{\substack{j,k,l=1\\k\neq (j,l)}}^N\frac{\left(3(\eb_{jk}\cdot\eb_{kl})^2-1\right)\eb_{kj}}{d_{jk}^2d_{kl}^3},
\label{U_free}
\end{align}
with $d_{jk}=|\Rb_j-\Rb_k|$ the interparticle distance and $\eb_{jk}=(\Rb_k-\Rb_j)/d_{jk}$ the unit vector directed from particle $j$ to particle $k$. The velocity of the center of mass is then
\begin{equation}
\Ub_{CM}=\frac{1}{N}\sum_{j=1}^N\Ub_j^\textrm{free}=\frac{5AM}{2N}\sum_{\substack{j,k,l=1\\ k\neq (j,l)}}^N\frac{\left(3(\eb_{jk}\cdot\eb_{kl})^2-1\right)\eb_{kj}}{d_{jk}^2d_{kl}^3}\cdot
\end{equation}

Considering now the final clustered configuration (Phase II), the translation and rotation velocities of the cluster, $\Ub_\textrm{cluster}$ and $\Omegab_\textrm{cluster}$ are obtained as
\begin{align}
\Ub_\textrm{cluster} + \Omegab_\textrm{cluster} \times \Rb_j = & \mathbf{U}^\textrm{free}_j + \frac{\mathbf{F}_j}{6\pi}  +\sum_{k\neq j} \left[\frac{\mathbf{I} +\eb_{jk}\eb_{jk}}{8\pi d_{jk}}-\frac{3\eb_{jk}\eb_{jk}-\mathbf{I}}{12\pi d_{jk}^3}\right]\cdot \mathbf{F}_k\qquad \textrm{with   }\sum_{j}\Fb_j  =0. 
\label{eq:mor_cluster} 
\end{align}

These results are valid up to $O(\varepsilon^6)$ corrections with $\varepsilon\sim a/d$ the typical radius-to-distance ratio.
At leading order in  $\varepsilon$, these results indeed confirm that
\renewcommand{\theenumi}{\roman{enumi}}%
\begin{enumerate}
\item in the clustering phase (Phase I), particles exhibit $O(\varepsilon^2)$ individual velocities, while their center of mass velocity is $O(\varepsilon^5)$,
\item in the self-propulsion phase (Phase II), the global cluster velocity is $O(\varepsilon^3)$ and there is no relative motion.
\end{enumerate}

The analysis presented in Appendix~\ref{app:mor} also demonstrates the physical origin of this increased global motion as a consequence of the non-uniform hydrodynamic resistance experienced by the different particles within the cluster, illustrated by the right-hand-side of Eq.~\eqref{eq:mor_cluster}. At leading order, the phoretic forcing is pairwise and symmetric and therefore averages to zero. Differences in hydrodynamic resistance between two particles however introduces an imbalance that must be compensated by a global translation and/or rotation.

This decomposition of the dynamics of the particles into two successive and physically distinct phases is of further importance. Self-propulsion is essentially restricted to Phase II where the particles are clustered together. But, since the self-propulsion of the particles depend on the geometry of the cluster formed as an outcome of the clustering phase (Phase I), the self-propulsion velocity of the assembly is indirectly, but entirely determined by the detailed kinematic process of phoretic clustering.

\subsection{A reduced-order model of the clustering dynamics: pairwise interactions}
\label{subsec:pair}

Characterizing the collective self-propulsion properties of isotropic 
particles therefore critically relies on a good understanding and modelling of the clustering phase. The clustering process conditions the final shape and therefore collective propulsion properties. Besides the maximum velocity achievable for $N$ particles, prominent self-propulsion properties such as the mean and most probable velocities require determining the probability of formation of a given cluster configuration through the phoretic attraction of $N$ particles that are initially randomly-distributed. The approach chosen here is a Monte-Carlo framework where, for each value of $N$, a large number of clustering simulations are run starting from random initial particle arrangements, and the probability $\mathcal{P}_q^\textrm{phoretic}$ to obtain a given cluster shape (indexed by $q$) is computed from the number of runs leading to that particular cluster shape. Due to their attractive nature, phoretic interactions of the particles tend to maximize their packing in their clustered configuration. Restricting here our analysis to two-dimensional clusters, particles are arranged on a regular hexagonal lattice in their final configuration. For $N\leq 5$ particles, a single stable shape is obtained (see Figure~\ref{fig:5pmesh}), while for $N>5$, the number of final distinct cluster configurations is finite but increases exponentially with $N$; for example: $N=6$, $N=8$ and $N=10$ lead to $3$, $9$ and $35$ distinct configurations, respectively. 

Boundary Element Methods are well suited to compute the detailed dynamics accurately, but are prohibitively expensive for running thousands of simulations  of the full temporal dynamics as needed for obtaining the probabilities of different cluster shapes starting from random initial positions (see section \ref{subsec:BEM}). Motivated by the distinct features of the clustering and propelling phases identified in the previous section, our approach is therefore to split the problem into two distinct parts: (i) determine the distribution of probabilities for the different cluster shapes using a reduced-order model of the clustering dynamics, and (ii) compute the exact propulsion velocity of each final cluster using regularized BEM.

The reduced-order model of the clustering phase required for the first part must be sufficiently accurate, both in the far-field and near-field limits, so that the final configuration is the same as that predicted by BEM, yet be sufficiently inexpensive computationally to be able to run a large number of simulations for each $N$. At leading order (i.e. far-field approximation), the relative velocity of the particles is determined as the superposition of symmetric interactions between pairs of particles. However, this method fails to provide accurate solutions when the separations between particles are small ($d_c<a$). The method of reflections detailed in Appendix~\ref{app:mor}, which provides iterative approximations of increasing accuracy, is an appealing alternative to full simulations as it can capture the multi-particle dynamics, but  it is fundamentally restricted to distances greater than the size of the particle ($a/d<1$) for numbers of reflections small enough for practical implementation.

Turning back to the two-particle case, an exact solution for the Laplace and Stokes problems was obtained in Section~\ref{sec:2p} which includes a full description of confinement and lubrication effects. We exploit this solution here to account for far- and near-field dynamics properly, by superimposing the pairwise phoretic drift velocities between the particles whilst retaining the full exact solution in Eq.~\eqref{eq:complete} $U_a(d_{jk})$ (including the repulsive interaction that prevents particle overlap). Note that a fundamental restriction of this superposition assumption is its inability to predict any collective propulsion: the velocity of the center of mass of the arrangement is identically zero by definition. Yet, we observed in the full dynamics that such center of mass motion is negligible during the clustering phase, so that this constraint has little physical implication in practice. It should further be emphasized that this superposition of pairwise solutions is not derived from first principles, but rather on its practical ability to capture correctly the leading order interactions both for near- and far-field limits, and its practical ability to match the BEM predictions with good accuracy. 

\begin{figure}[t]
\begin{center}
\includegraphics[height=8cm]{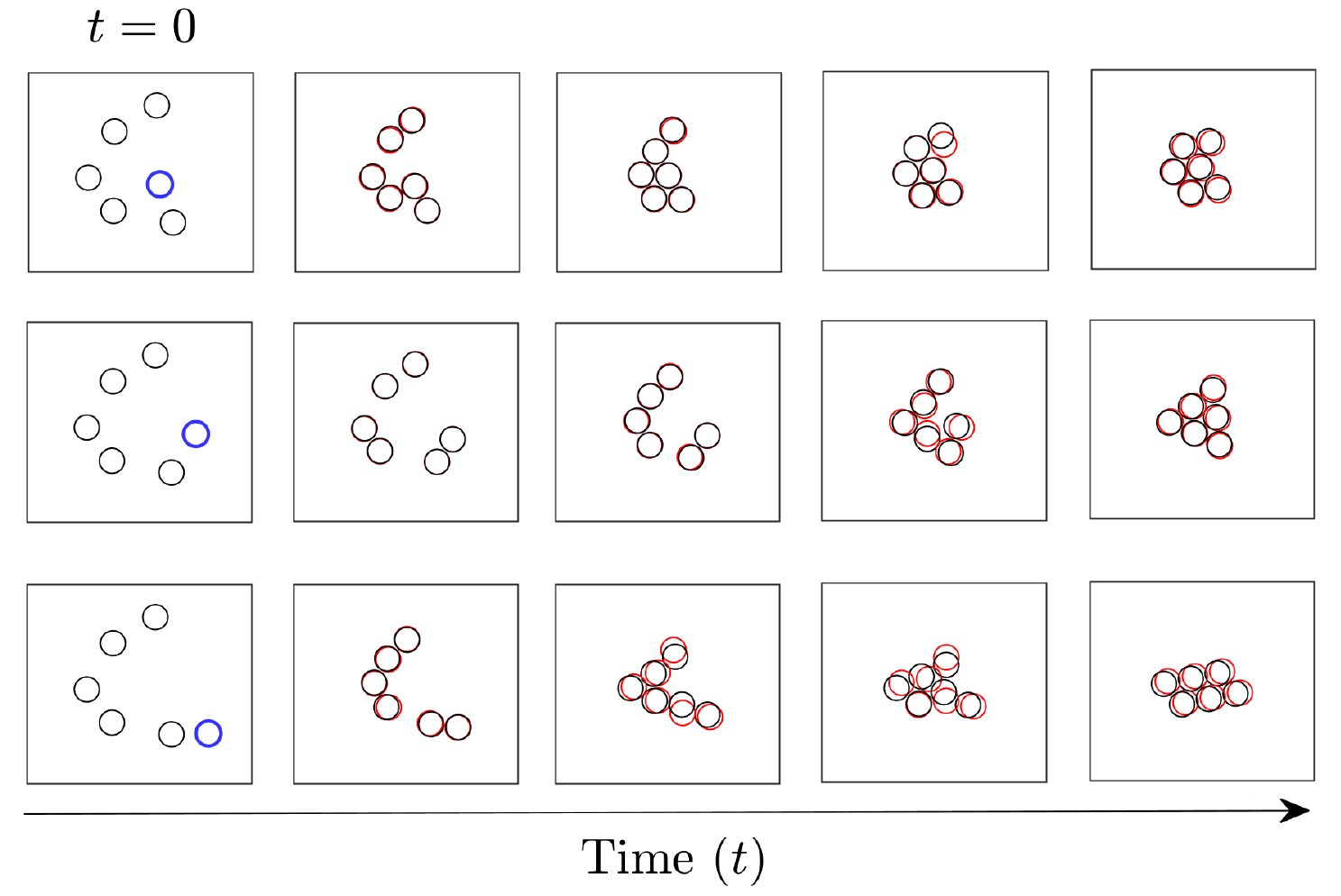}
\caption{Comparison of the positions of the particles predicted by BEM (in black) and reduced model (in red) at various instances. The three different initial distributions of the particles, which differ only in the position of a single particle (in blue), lead to three different cluster shape using BEM, all of which are correctly captured by the reduced-order model.}
\label{fig:BEM_qual}
\end{center}
\end{figure}

The accuracy of this reduced-order interaction model is therefore checked in details against BEM simulations in for $N=3$ and $N=6$ particles. 
\begin{figure}[t]
\begin{center}
\begin{tabular}{c}
\includegraphics[width=.75\textwidth]{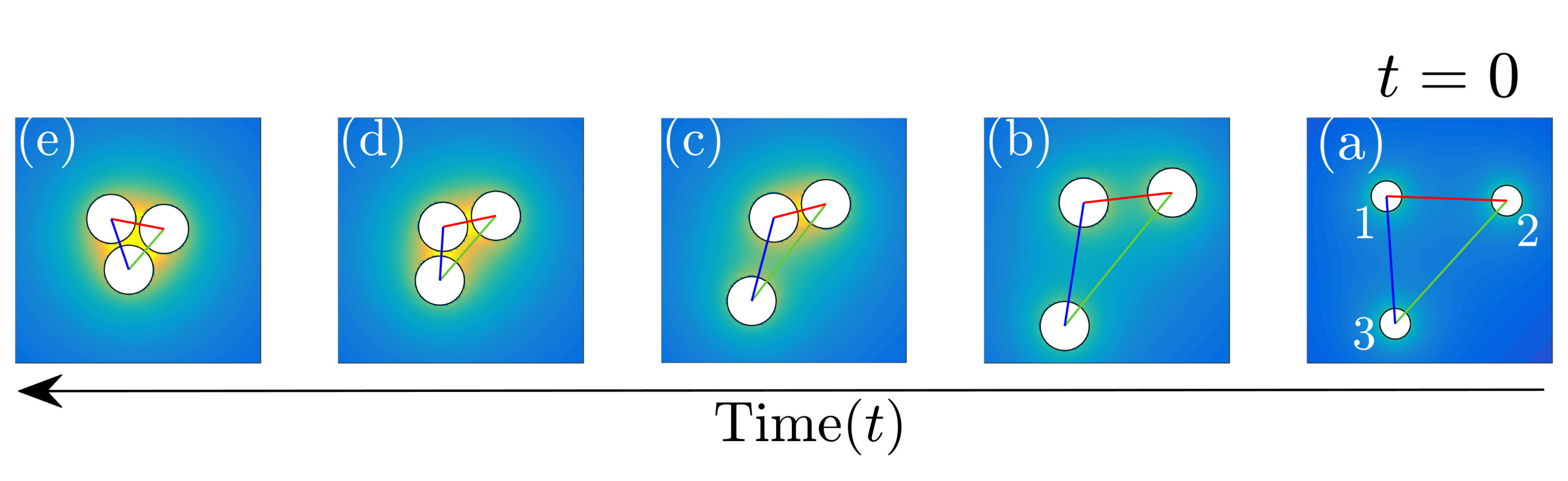}\\ \\
\includegraphics[width=.7\textwidth]{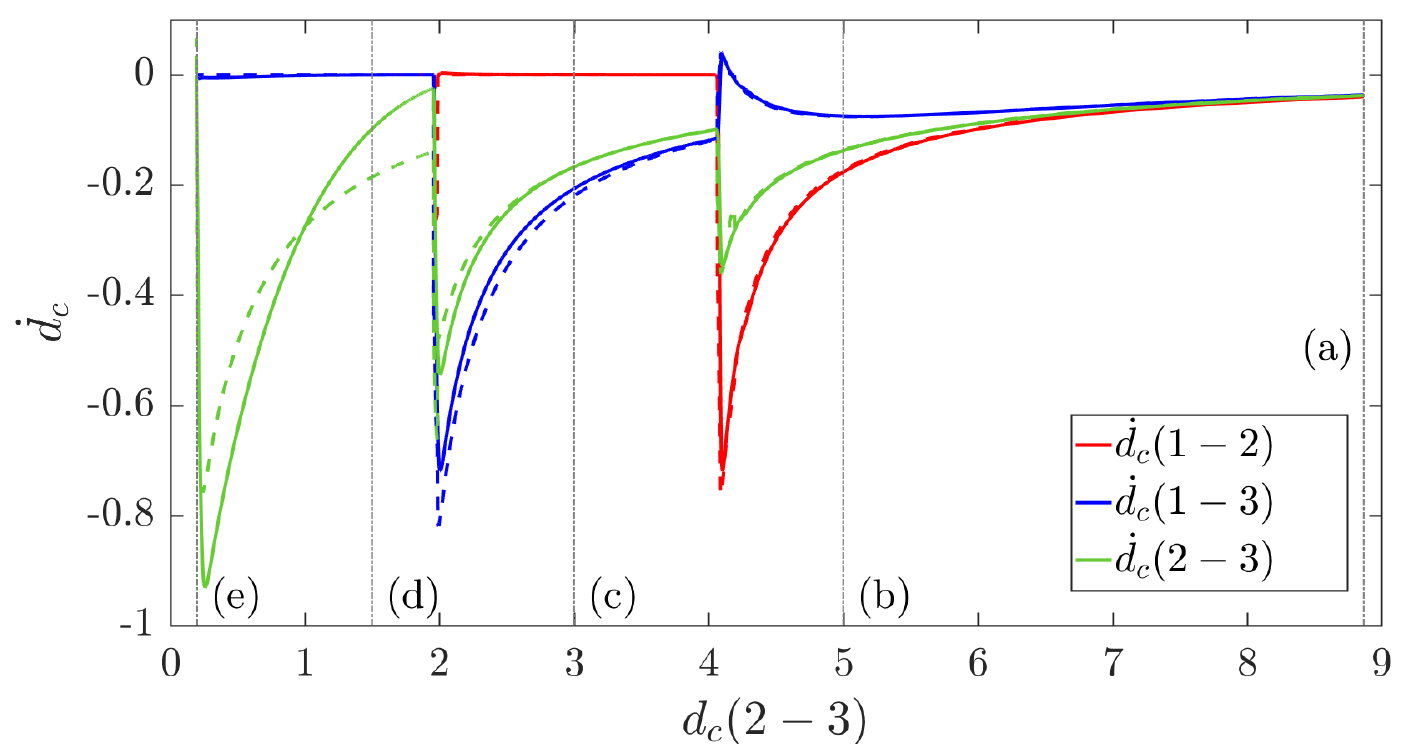}
\end{tabular}
\caption{Top: from right to left, snapshots of the phoretic clustering in a system of $N=3$ particles obtained using BEM. For visualization purposes, $(a)$ is zoomed out; the coloured lines represent the distance between centers of different pairs of particles. Bottom: The rate of change of separation ($\dot{d_c})$ between particles due to phoretic attraction predicted by the reduced model (dashed lines) and BEM (solid lines). The positions of the particles at each instant are set to that of the full BEM simulations. To avoid spurious repulsive velocities arising from slight differences in the equilibrium distance in the reduced model and BEM results and the stiffness of the repulsive model, the repulsive velocities in the reduced-order model are removed once two particles are in contact. The stages of clustering $(a)$ to $(e)$ are also shown (dashed-dotted lines). }
\label{fig:BEM_quant}
\end{center}
\end{figure} 
Figure~\ref{fig:BEM_quant} illustrates the clustering dynamics of $N=3$ particles starting from arbitrary initial conditions and quantitatively compares the rate of change in relative distance at each time as predicted by the reduced-order model to that obtained using full BEM simulations. The model is observed to predict the clustering velocities with excellent accuracy except for the final stages of clustering where particles are already in contact but for two of them, and the final shape is already fully determined. A second validation of the ability of the reduced-order model to predict the final configuration correctly is proposed in Figure~\ref{fig:BEM_qual} where  the evolution in time of three different initial arrangements of six particles ($N=6$) are plotted.

 These results confirm that this model is sufficient to capture the final shape of the cluster (which is the only piece of information that it will be used for): slight discrepancies in absolute positions of the particles arise only in the final stages of clustering, and do not affect their relative arrangement (i.e. they only correspond to a slight translation or rotation of the whole assembly). Hence, the reduced-order model is chosen in the following as a suitable substitute for BEM for modelling the clustering phase (Phase I) as it presents a good compromise between the accuracy of its prediction of the final shape and efficient computations (a few seconds are necessary for a single run, compared to tens of hours with BEM).

In practice, the velocity of particle $j$ is computed in this reduced-order model as
\begin{equation}
\totd{\Rb_j}{t}=\sum_{k\neq j}U_a(d_{jk})\eb_{jk},
\end{equation}
with $U_a(d)$ given in Eq.~\eqref{eq:complete}. One should note that this reduced model is formally equivalent to the motion of $N$ particles down the steepest gradient of an interaction potential $\mathcal{E}$:
\begin{equation}
\totd{\Rb_j}{t}=-\pard{\mathcal{E}}{\Rb_j}, 
\end{equation}
with
\begin{equation}
\mathcal{E}(\Rb_1,...,\Rb_n)=\sum_{j<k}\mathcal{E}_{2p}(|\Rb_j-\Rb_k|),\quad \textrm{and     }\pard{\mathcal{E}_{2p}}{d}=-U_a(d).
\label{eq:potentialN}
\end{equation}

It should be stressed here that this formal equivalence only holds rigorously for the reduced-order model, and not for the full problem (e.g. in the latter, phoretic interactions lead to a net global motion). Within the reduced-order model framework, particles arrange to minimize $\mathcal{E}$ which is equal to zero for infinitely-distant particles. As such, $\mathcal{E}$ can be understood as a measure of the cohesive interactions within and stability of a  particular cluster shape. Noting $\mathcal{E}_q$ its value for cluster shape $q$, a Boltzmann distribution would therefore be expected if the cluster formation was purely stochastic and driven solely by the stability of the final shape, with the probability $\mathcal{P}^\sigma_q$ to obtain cluster $q$ equal to
\begin{equation}
\mathcal{P}^\sigma_q=\frac{\ee^{-\mathcal{E}_q/(2\sigma^2)}}{\displaystyle\sum_m\ee^{-\mathcal{E}_m/(2\sigma^2)}},
\label{eq:P_noise}
\end{equation}
and $\sigma^2$ characterizes the background noise in the system: $\sigma=0$ leads to $\mathcal{P}^0_q=1$ for the most stable cluster (i.e. that with maximum $\mathcal{E}_q$), and $\mathcal{P}^0_q=0$ for all others, while $\sigma=\infty$ results in all cluster shapes having the same probability.

\section{Phoretic clustering and self-propulsion}
\label{sec:results}
\begin{figure}[t]
\begin{center}
\begin{tabular}{c}
\includegraphics[height=8.5cm]{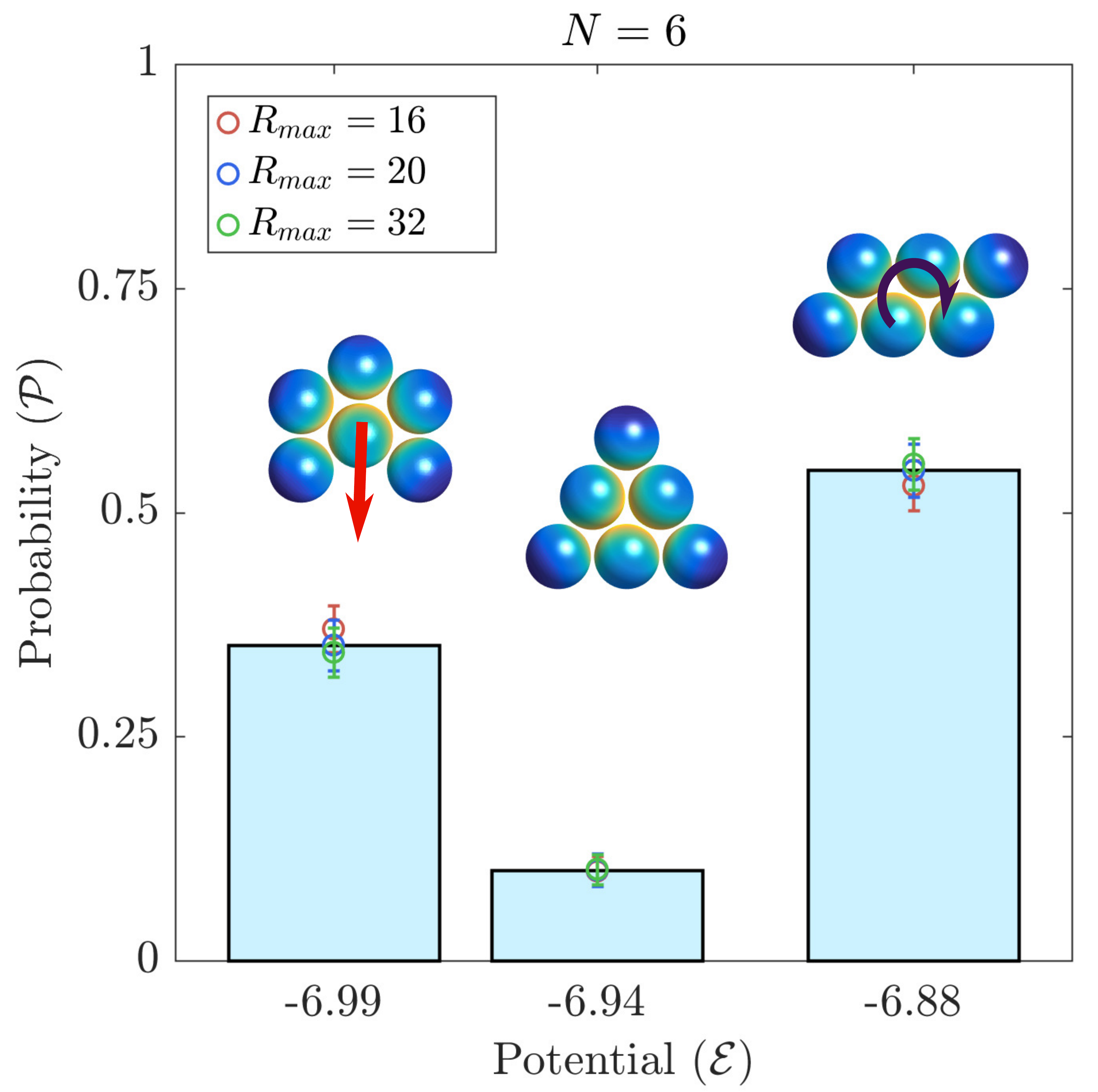}\\ \\
\includegraphics[height=8.5cm]{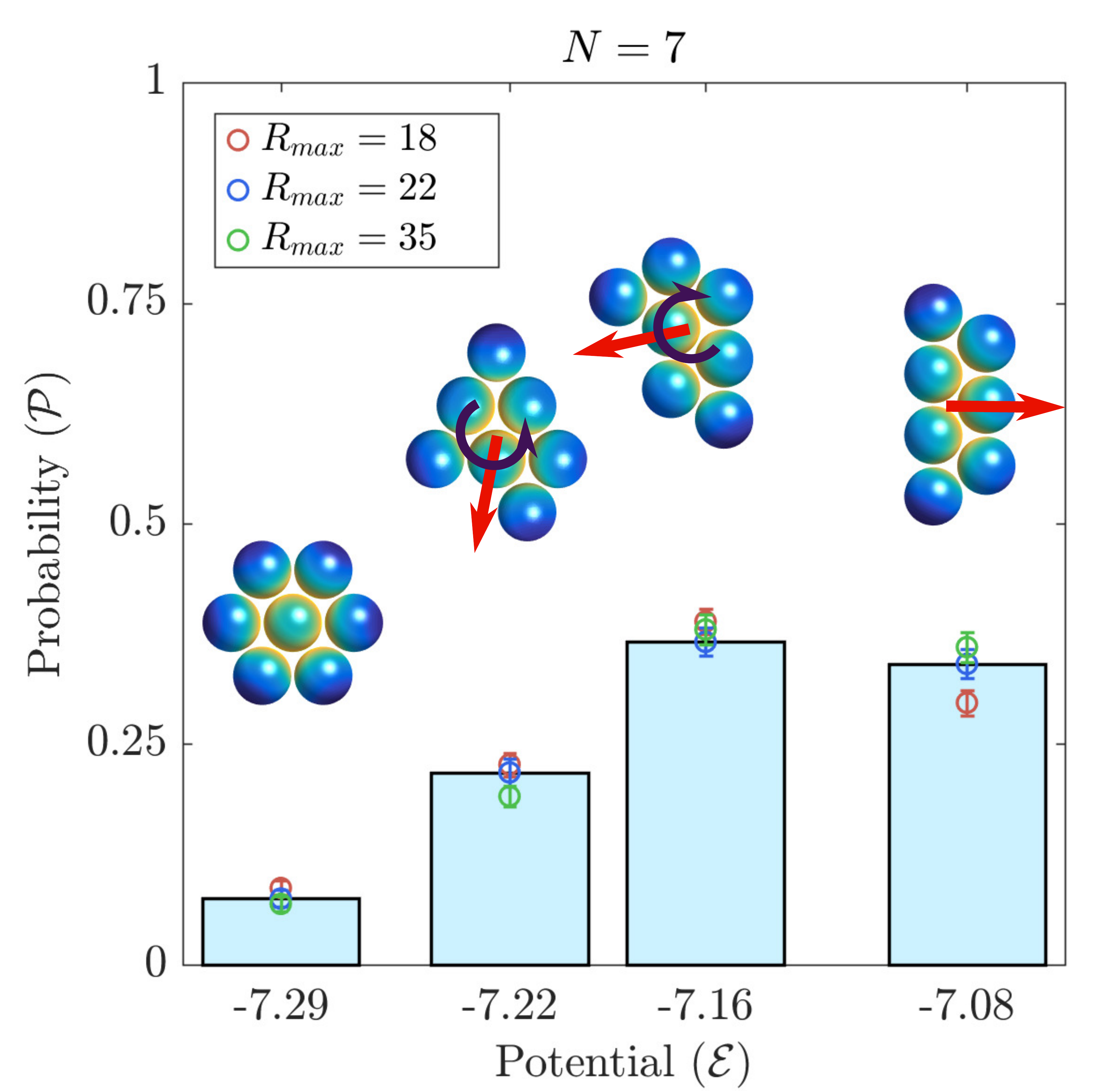}
\end{tabular}
\caption{Stable 6-particle and 7-particle clusters and their probabilities obtained from 2000 independent trials. Bar graph shows probabilities when particles begin at different random locations within a circle of radius $R_{max}$ for each trial (Histogram shown for $R_{max}=20 $ for 6-particle and $R_{max}=22$ for 7-particle system respectively). Measurements are made for different $R_{max}$ to show its influence. Range bars represent 2 standard deviation limits of probabilities computed from 50 such events.}
\label{prob_configs}
\end{center}
\end{figure}

Using the reduced-order clustering model described in the previous section (Phase I), a large number of simulations  are performed, with random initial spatial arrangements of particles in order to obtain the probability of formation of the different cluster configurations through phoretic attractions. In a second step, the self-propulsion velocity of each cluster (Phase II) is computed accurately using BEM and hence, the self-propulsion statistics are obtained for a given system of $N$ particles. (i.e. maximum, mean and most probable velocities)  

\subsection{Probabilities of formation of stable rigid clusters}
\label{subsec:prob}

To study the complete evolution of clustering of an $N$-particle system from zero interaction potential to a final cluster potential $\mathcal{E}_q$, the particles have to ideally begin from an infinite separation. For the practical purpose of simulations, the particles are initially distributed randomly within a large disk of radius $R_{max}$. It is ensured that $R_{max}$ is sufficiently large that it does not significantly affect the probability statistics (Figure~\ref{prob_configs}). For $N=6$, which is the smallest value of $N$ for which multiple stable configurations are obtained, probability values converge for $R_{max} \gtrsim 20$ (see Figure~\ref{prob_configs}), which corresponds to a density (area fraction of particles in the clustering plane) of $1.5\%$. For all $N \leq 12$, it is observed that this area fraction of $1.5\%$ is sufficient to give accurate probability statistics.

The number of independent trials needed to determine the probabilities of configurations accurately increases with $N$. We found that for $N=6$ and $N=7$, 2000 distinct runs are sufficient and the resulting probability distributions and accuracy are shown in Figure~\ref{prob_configs}. Clusters are labelled by their effective interaction potential $\mathcal{E}_q$, i.e. the most stable cluster is on the left. 

Interestingly, Figure~\ref{prob_configs} shows that the most stable cluster (i.e. that with least effective potential $\mathcal{E}_q$ in the reduced-order model) does not  have the highest probability as one would expect in a stochastic process. The phoretic clustering described by Eq.~\eqref{eq:potentialN} (i.e. down-gradient of the interaction potential) does not lead to an absolute but to a local minimum, as it depends on the detailed route followed in the configuration phase space during the clustering process. Less ``stable'' cluster shapes minimize the interaction potential only locally but may be wider attractors in the ($2N-3$)-dimensional phase space that characterizes the clustering motion. Because of the intimate link of cluster shape and velocity, this is expected to hold profound consequences on the collective self-propulsion properties of the $N$ particles.

\begin{figure}[t]
\begin{center}
\includegraphics[width=.55\textwidth]{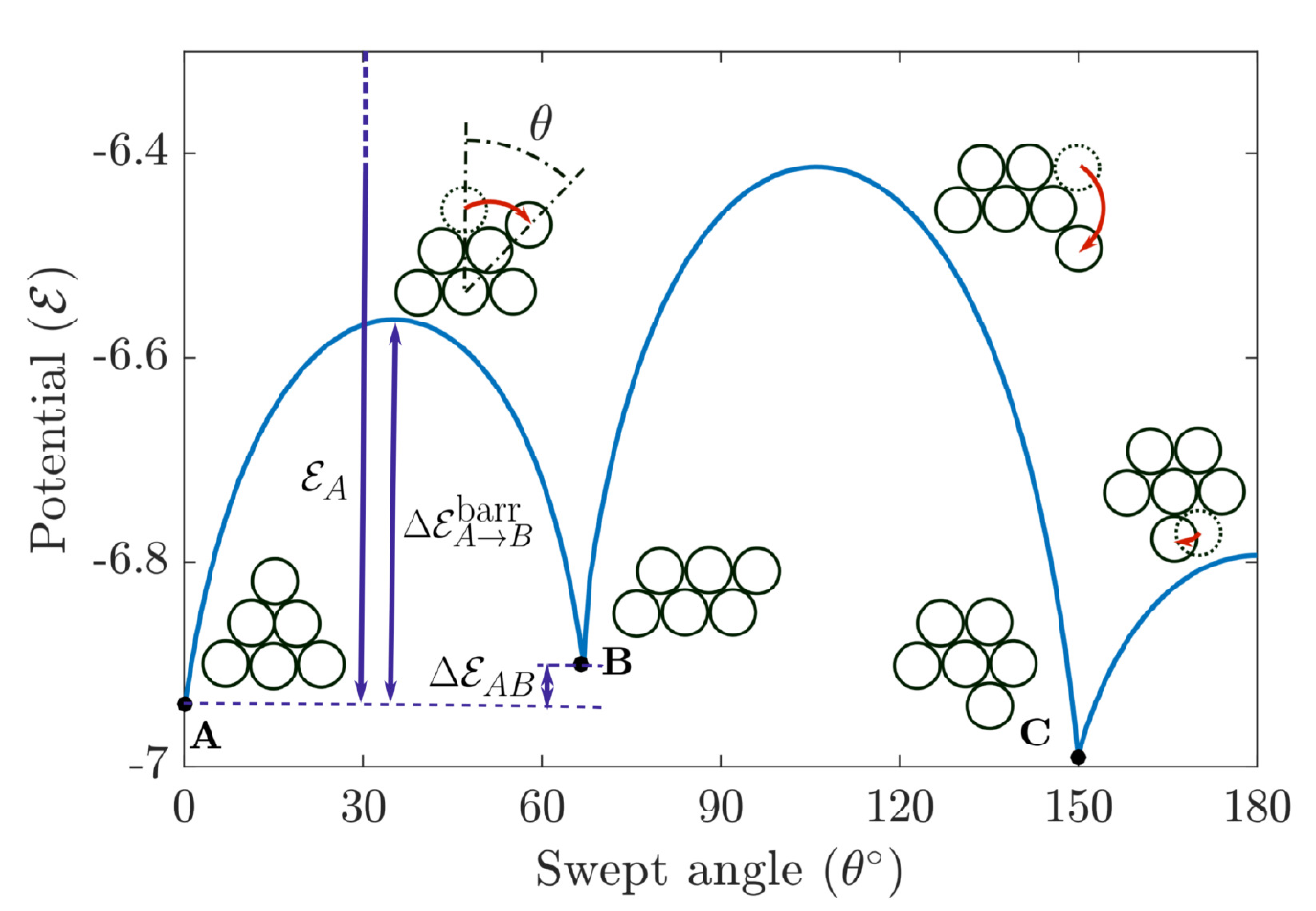}
\caption{Evolution of the assembly potential while rolling a single particle around the cluster from an initial stable configuration $A$ of potential $\mathcal{E}_A = -6.94$ in order to sample all configurations. Clusters $A$, $B$ and $C$ represent local minima of the effective interaction potential $\mathcal{E}$. $\Delta \mathcal{E}_{A\to B}^{\scriptsize \mbox{barr}} \approx 0.4 $ represents the minimum gain in effective potential required to transition from cluster $A$ to cluster $B$, while the actual potential difference between clusters is just $\Delta \mathcal{E}_{AB} \approx 0.05$. $|\mathcal{E}_A|\approx 7$ is the change in effective potential between the initial stage and final clustered configuration. The results for $180^\circ\leq \theta\leq 360^\circ$ are obtained by symmetry. }
\label{fig:roll}
\end{center}
\end{figure}

Once formed, a cluster cannot transition to another configuration without additional forcing (i.e. ``energy'' input). Even though the differences in the final potential between configurations $(\Delta \mathcal{E}_{pq} = |\mathcal{E}_p-\mathcal{E}_q|)$ are relatively small in comparison to $|\mathcal{E}_q|$ and $|\mathcal{E}_p|$ (i.e. the change in potential from the initially-dispersed configuration to the final clustered shape), switching from one configuration to another requires overcoming a much larger potential barrier ($\Delta \mathcal{E}_{p\to q}^{\mbox{\scriptsize barr}}$). This is illustrated in Figure~\ref{fig:roll} in the case of $N=6$ particles: changing the position of a single particle around the rest of the cluster allows the arrangement to describe all three possible stable configurations (identified as minima in effective potential $\mathcal{E}$). 

\begin{figure*}[t]
\begin{center}
\includegraphics[width=.92\textwidth]{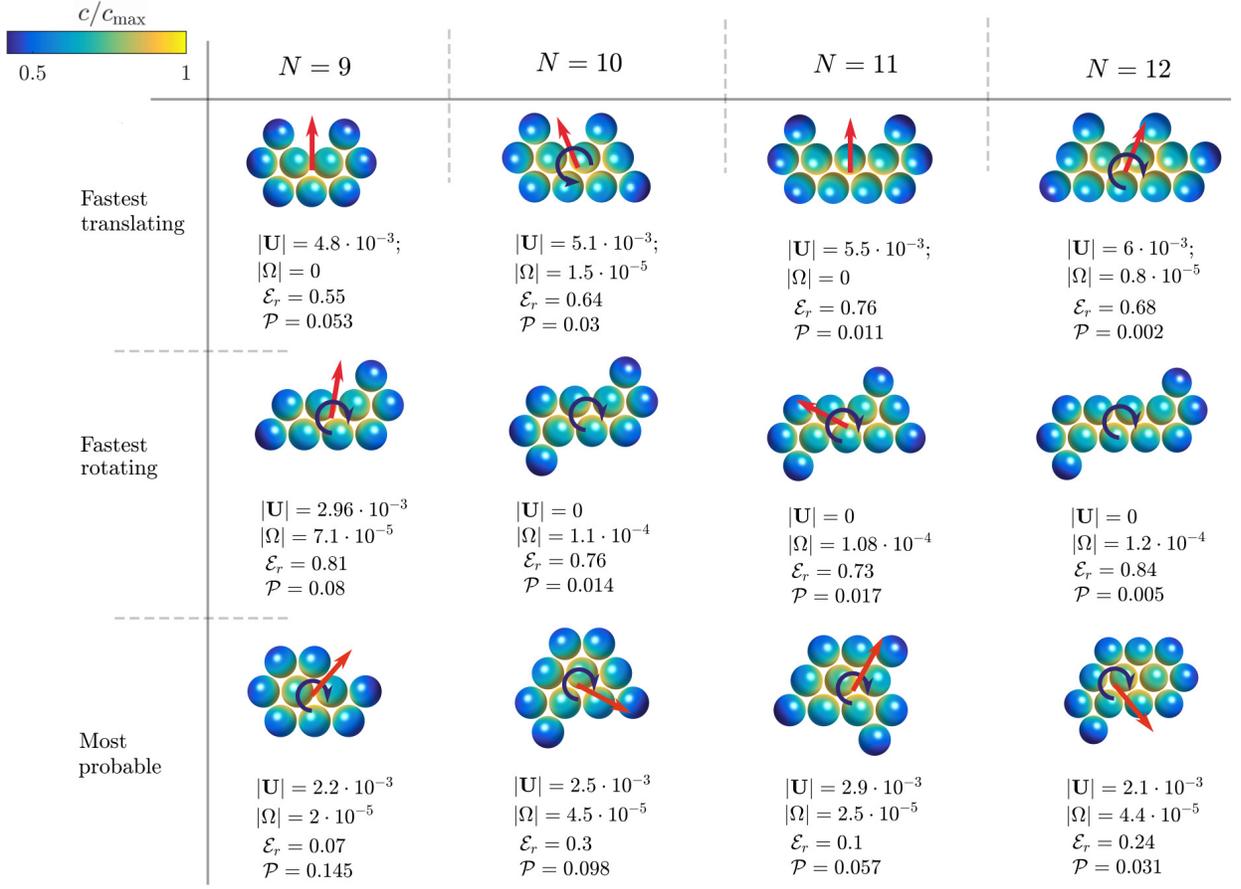}
\caption {Fastest translating (top), fastest rotating (center) and most probable clusters (bottom) for $N=9$ to $12$. The relative surface concentration is shown in color for each cluster as well as its propulsion, stability and probability characteristics.}
\label{fastest}
\end{center}
\end{figure*}

While the difference in effective potential of the three stable configurations is small $\Delta \mathcal{E}_{pq}\sim O(0.01)$ ($\{p,q\} \subset \{A,B,C\}$) , transition from one configuration to another requires reaching intermediate stages representing a potential barrier typically an order of magnitude larger, $\Delta \mathcal{E}_{p\to q}^{\scriptsize \mbox{barr}}\sim O(0.1)$. Note also that the difference in effective potential between the pre-clustering state (ideally, $\mathcal{E}=0$) and final configuration is yet an order of magnitude larger $|\mathcal{E}_q| \sim O(1)$.

\subsection{Propulsion velocities of rigid clusters}

Geometrically-asymmetric phoretic systems self-propel with a global translation and/or rotation velocity determined only by the asymmetry of their geometric shape~\cite{michelin2015a,shklyaev2015}; in particular, this velocity is independent of the particles' size. As expected from symmetry arguments, configurations with a single mirror plane of symmetry cannot undergo any rotation, while configurations which are antisymmetric about a plane (left unchanged by a $180^{\circ}$ rotation) do not translate. Moreover, configurations with rotational symmetry neither translate nor rotate (see Figure~\ref{prob_configs} for example). 

As explained, in this clustered arrangement, the relative positions of the particles are fixed in a regular hexagonal lattice. For each $N$-particle system, the regularised Boundary Element Method detailed in Section~\ref{subsec:BEM} is used to compute the  exact translational and angular velocities of propulsion (Phase II) of all possible cluster configurations. A list of the fastest propelling and rotating clusters, as well as the most probable ones are shown in Figure~\ref{fastest} for $9\leq N\leq 12$. By defining the normalised effective potential for shape $q$ for fixed $N$ as $\mathcal{E}_r =(\mathcal{E}_q - \mathcal{E}_{\textrm{min}})/(\mathcal{E}_{\textrm{max}} - \mathcal{E}_{\textrm{min}})$, we observe that the fastest propelling (translation or rotation) clusters are some of the least stable clusters with respect to the large effective potential $\mathcal{E}_r$. This is the result of their large geometric eccentricity, responsible for their larger propulsion velocity through the larger concentration gradients at their surface it creates. In contrast, the most probable clusters are characterized by their more compact and roughly symmetric arrangement around their geometric center (see Figure~\ref{fastest}). 

\subsection{Clustering-induced self-propulsion properties}
\label{subsec:prop}

\begin{figure}[t]
\begin{center}
\begin{tabular}{c}
\includegraphics[height=6.5cm]{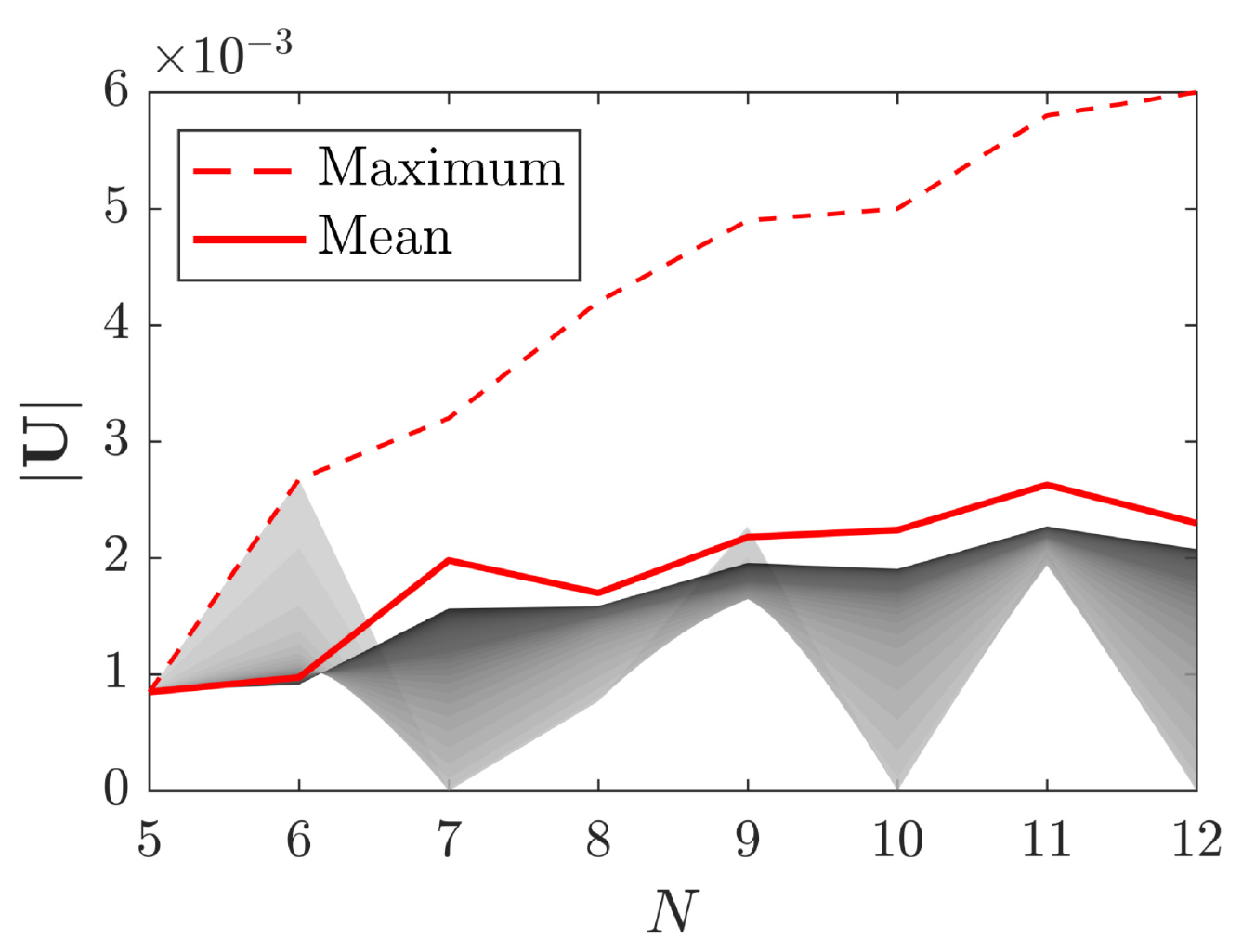} \\
\includegraphics[height=6.5cm]{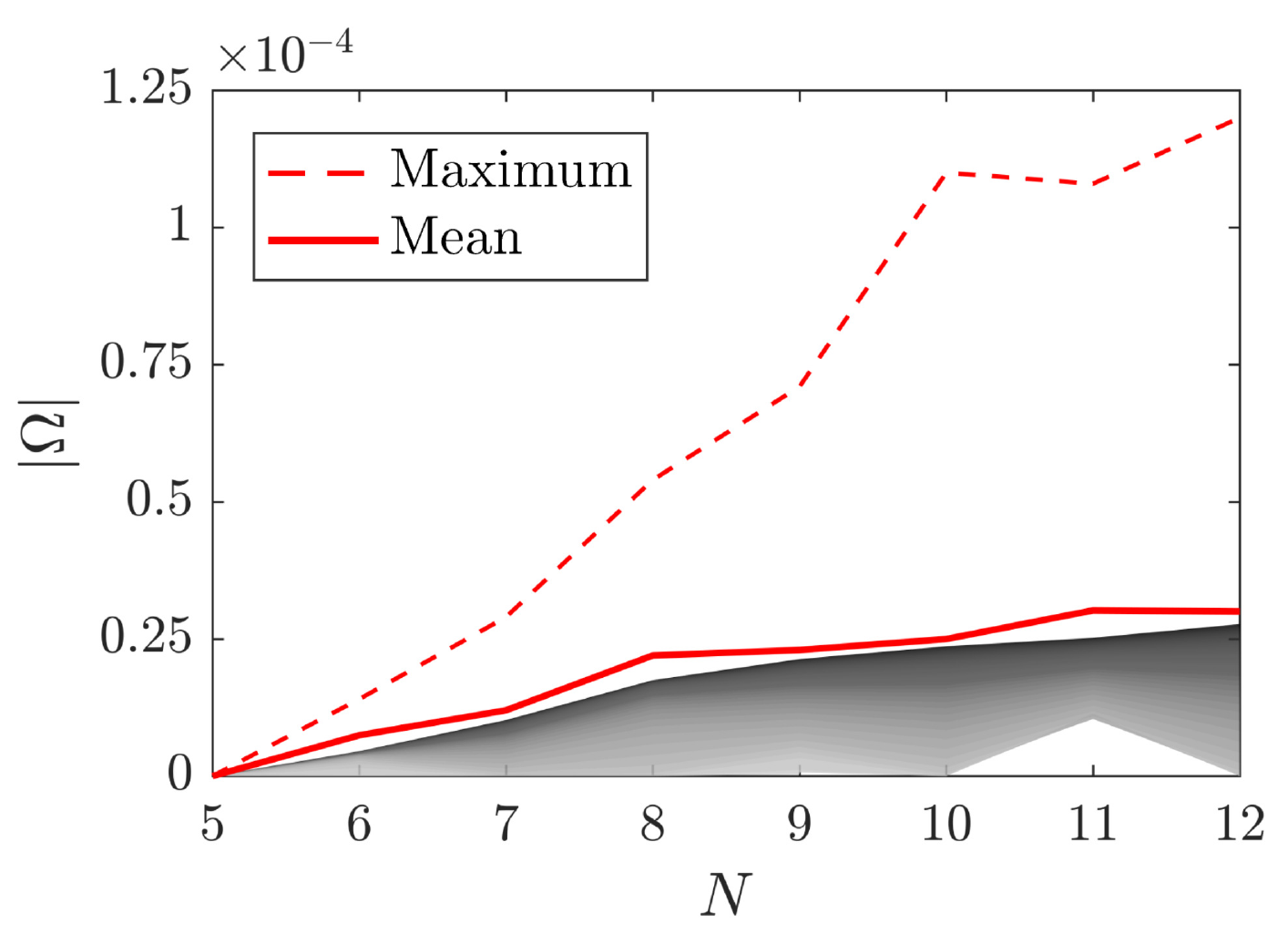}
\end{tabular}
\caption{Magnitudes of maximum (dashed) and mean (solid) translational and rotational velocities of clusters obtained using BEM. The grey shading represents the mean velocity obtained through an equilibrium distribution, Eq.~\eqref{eq:P_noise} with the noise level $\sigma^2$, increasing from zero to infinity as the intensity of shading gets darker.}
\label{meanmax}
\end{center}
\end{figure}

Combining the results of the previous sections, the statistical properties of the collective self-propulsion of $N$ isotropic particles resulting from their phoretic clustering are obtained here. Figure~\ref{meanmax} shows the mean and maximum velocities of $N$-particle clusters. In particular, for a given $N$, noting $\mathcal{P}_q$ the probability and $U_q$ the velocity magnitude (or equivalently, rotational velocity magnitude $\Omega_q$) of cluster shape $q$, the mean velocity is defined as
\begin{equation}
U_\textrm{mean}(N)=\sum_q\mathcal{P}_q U_q,\quad \Omega_\textrm{mean}(N)=\sum_q\mathcal{P}_q \Omega_q,
\end{equation}
with the sum carried out on all possible cluster shapes.

\begin{figure*}[t]
\begin{center}
\begin{tabular}{c}
\includegraphics[height=4cm]{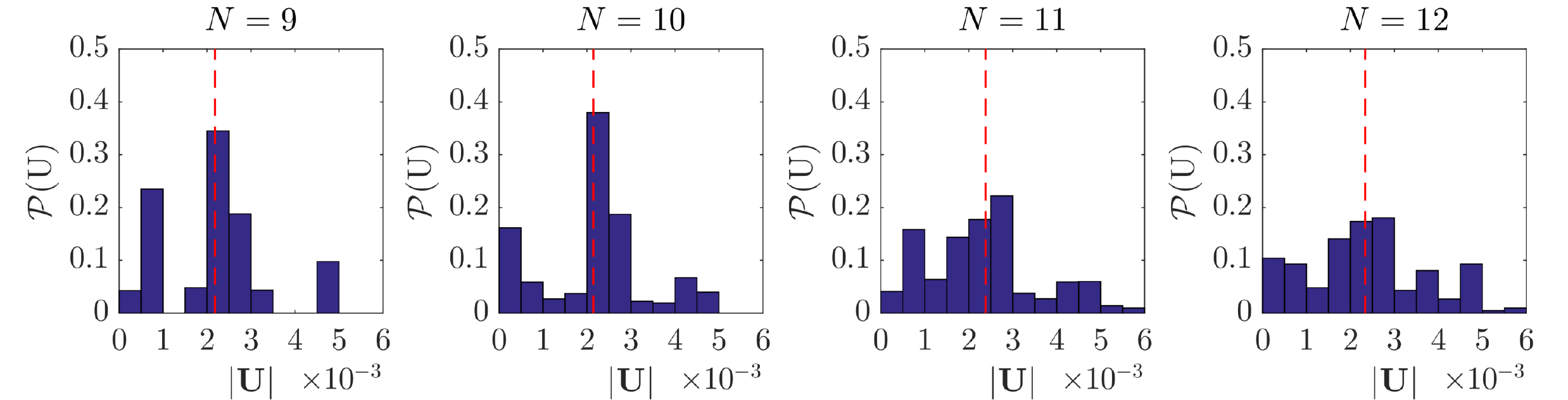}\\
\\
\includegraphics[height=4cm]{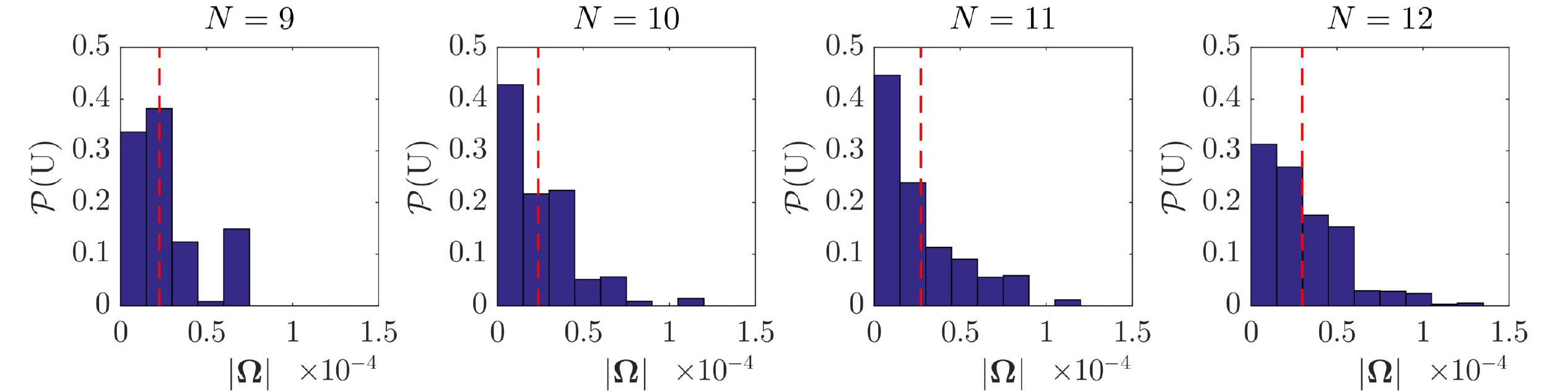}
\end{tabular}
\caption{Probability distribution of (top) translational velocities \textbf{U} and (bottom) rotational velocities $\Omegab$ of $N$-particle clusters. The distribution mean is indicated by a dashed red line for each $N$.}\label{histo_vel_rot}
\end{center}
\end{figure*}

Because of the symmetry of the only existing final cluster, no self-propulsion is observed for $N\leq 4$. Note that for all $N>5$, at least one non-propelling cluster (having rotational symmetry) is found; and therefore the minimum velocity for all $N>5$ is strictly zero. The maximum propulsion velocity is observed to increase steadily with $N$ as the larger number of particles allows for more eccentric shapes and larger phoretic forcing.  However, the  mean velocity $U_\textrm{mean}(N)$ is observed to saturate for larger values of $N$. Insight on this result as well as other statistical properties of importance such as the most probable velocity and the variance in propulsion velocities of large $N$-particle clusters is obtained by studying the probability distribution of the propulsion velocities (Figure~\ref{histo_vel_rot}).

The probability distributions of translational velocities of clusters with $N\geq 10$ show a prominent peak near the mean value of the distribution ($\approx 2 \times 10^{-3}$) indicating that clusters with this velocity are also the most likely to form. The most probable angular velocity, however, is lower than the mean value ($\approx 2.5 \times 10^{-5}$). As the maximum velocity (translation and rotation) increases with $N$, the graph spreads indicating an increase in variance of velocities between the various configurations. However, the fastest propelling configurations, which are also some of the most eccentric and elongated ones, have very low probability so that such clusters are only seldom observed for large $N$, and hence their contribution to the mean velocity is minimal. For $N\geq 10$, most of the contribution to the mean properties is brought by clusters with an intermediate velocity which remains relatively fixed with $N$, leading to the saturation in the mean velocity as $N$ grows. This can be qualitatively understood from the most dominant features of the most probable clusters as depicted in Figure~\ref{fastest}: those clusters display more compact shapes with an asymmetry arising only from a small number of particles.

It was already emphasized above that the probability of formation of a given shape in the phoretic clustering process does not correlate with its effective potential $\mathcal{E}$ for the reduced-order model, as would be expected in a classical system at thermodynamic equilibrium, for which the probability of formation of various configurations follows a Boltzmann distribution, Eq.~\eqref{eq:P_noise}. Not unexpectedly, this has consequences for the collective propulsion properties: Figure~\ref{meanmax} shows that the mean velocity resulting from phoretic interactions is systematically larger than it would be under the sole constraint of minimizing the effective interaction potential, regardless of the importance of noise in the process.

\section{Effect of noise}
\label{sec:noise}
The previous sections focused on the clustering-induced collective dynamics of phoretic particles, \emph{in purely deterministic systems}: for a given particles' arrangement at time $t$, the evolution of the particles' position in time is fully determined. Yet, in practice, all such systems which focus on microswimmers are subject to a variety of stochastic processes (including thermal noise) which continuously alter their motion. In the presence of background noise caused by temperature of the surrounding fluid, passive microscopic particles (or isolated isotropic phoretic swimmers) undergo Brownian motion characterized by zero mean displacement. Active particles however have a net displacement with continuous reorientation of the direction of propulsion, thus exhibiting diffusive behaviour in long time scales \cite{volpe2014,Bechinger}. Commonly employed minimal models for these \textit{Active Brownian Particles} describe the essential dynamics involving overdamped motion (Langevin dynamics) as well as their thermal reorientation \cite{Zottl, Bechinger}. 

Equivalently, in the case of isotropic phoretic particles, we describe the clustering process by its reduced-order model (see Section~\ref{subsec:pair}) rather than using the full BEM simulations because of its versatility and reduced computational cost, as well as to maintain consistency with the rest of the manuscript. Hence, the deterministic dynamics of the particles corresponds to a collective minimization of the effective interaction potential $\mathcal{E}$ as defined in Eq.~\eqref{eq:potentialN}. The purpose of this section is therefore to provide some insight on the robustness of these deterministic results, i.e. how the results of self-propulsion statistics obtained in the previous sections are modified in the presence of background noise on the kinematics of individual particles.

In the absence of inertia, the evolution of the position $\Rb_j^{(b)}(t)$ of particle $j$ under the effect of background noise is given by the overdamped Langevin equation:
\begin{equation}
\totd{\Rb_j^{(b)}(t)}{t} = \Ub_j(t) + \boldsymbol{\xi}_j(t),
\end{equation} 
where its deterministic velocity, $\Ub_j(t)$ is given by Eq.~\eqref{eq:potentialN}, and $\boldsymbol{\xi}_j(t)$ is a external Gaussian white noise, with zero mean and a variance, $\sigma^2 \mathbf{I}=<\boldsymbol{\xi}_i(t)\;\boldsymbol{\xi}_j(t')>=2D \delta(t-t')\delta_{ij}\mathbf{I}$, where $D$ is the diffusivity of each particle in the fluid. Thus, the instantaneous displacement of particle $j$ at any time $t$ is 
\begin{equation}
\mbox{d}\Rb_j^{(b)}(t) =  \Ub_j(t)\mbox{d}t+\mbox{d}\mathbf{W}_j,
\label{lang_d}
\end{equation}
where $\mathbf{W}_j$ is a Weiner process with zero mean and variance $ \sigma^2 t\; \delta_{ij} \mathbf{I}$. Discretizing Eq.~\eqref{lang_d} by using Euler-Mayurama method gives
\begin{equation}
\Rb_j^{(b)} (t+\Delta t) = \Rb_j^{(b)}(t)+  \Ub_j(t)\; \Delta t+\Delta\mathbf{W}_j,
\label{mayur}
\end{equation}
$\Delta \mathbf{W}_j$ is also a zero mean Weiner process with variance $ \sigma^2 \Delta t\; \delta_{ij} \mathbf{I}$. Equation~\eqref{mayur} is solved using an adaptive time stepping method. 

A system of $N=6$ particles is the smallest system that exhibits multiple clustered configurations, and for which the introduction of noise is expected to potentially introduce significant modification in the collective dynamics; we shall henceforth consider $N=6$ as an example to illustrate the effect of noise on the clustering statistics and resulting propulsion properties. A direct result of the component of randomness in position of particle in Eq.~\eqref{mayur}, is the formation of multiple configurations from the same initial conditions of the system. For each set of initial conditions (typically a few hundreds), Eq.~\eqref{mayur} is used to run about $100$ simulations. If the strength of noise satisfies $\sigma^2\sim|\mathcal{E}_q|$, background fluctuations are sufficient to break the formed clusters and redistribute the particles far apart. Hence, we restrict ourselves to the case where $\sigma^2$ is much smaller than the absolute cluster potential, $|\mathcal{E}_q|$ so that phoretic effects are still dominant and clustering occurs. In the following, two different behaviours are observed under the effect of noise, depending on its relative magnitude ($\sigma^2$) and the effective potential barrier $\Delta\mathcal{E}_{p\to q}^{\scriptsize \mbox{barr}}$ from configurations $p$ to $q$ (see Figure~\ref{fig:roll}). 

\subsection{Low noise}
\begin{figure}[t]
\begin{center}
\begin{tabular}{cc}
\includegraphics[height=6cm]{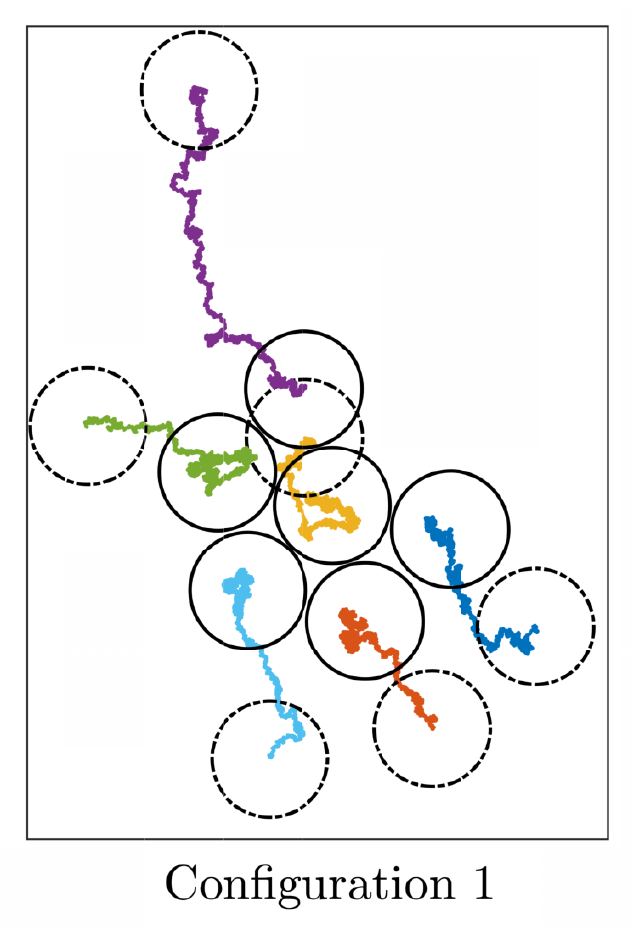} &
\includegraphics[height=6cm]{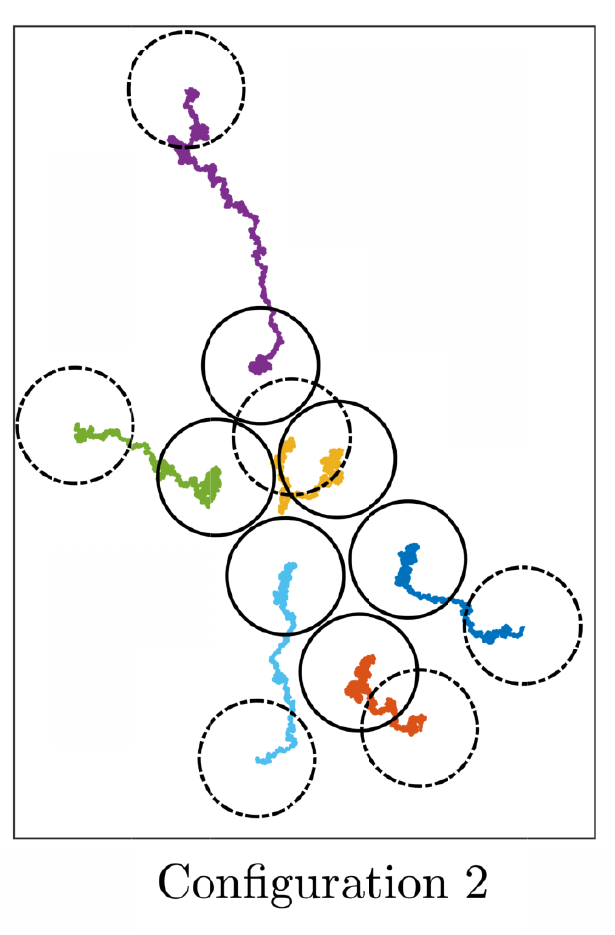} \\
\multicolumn{2}{c}{\includegraphics[height=6cm]{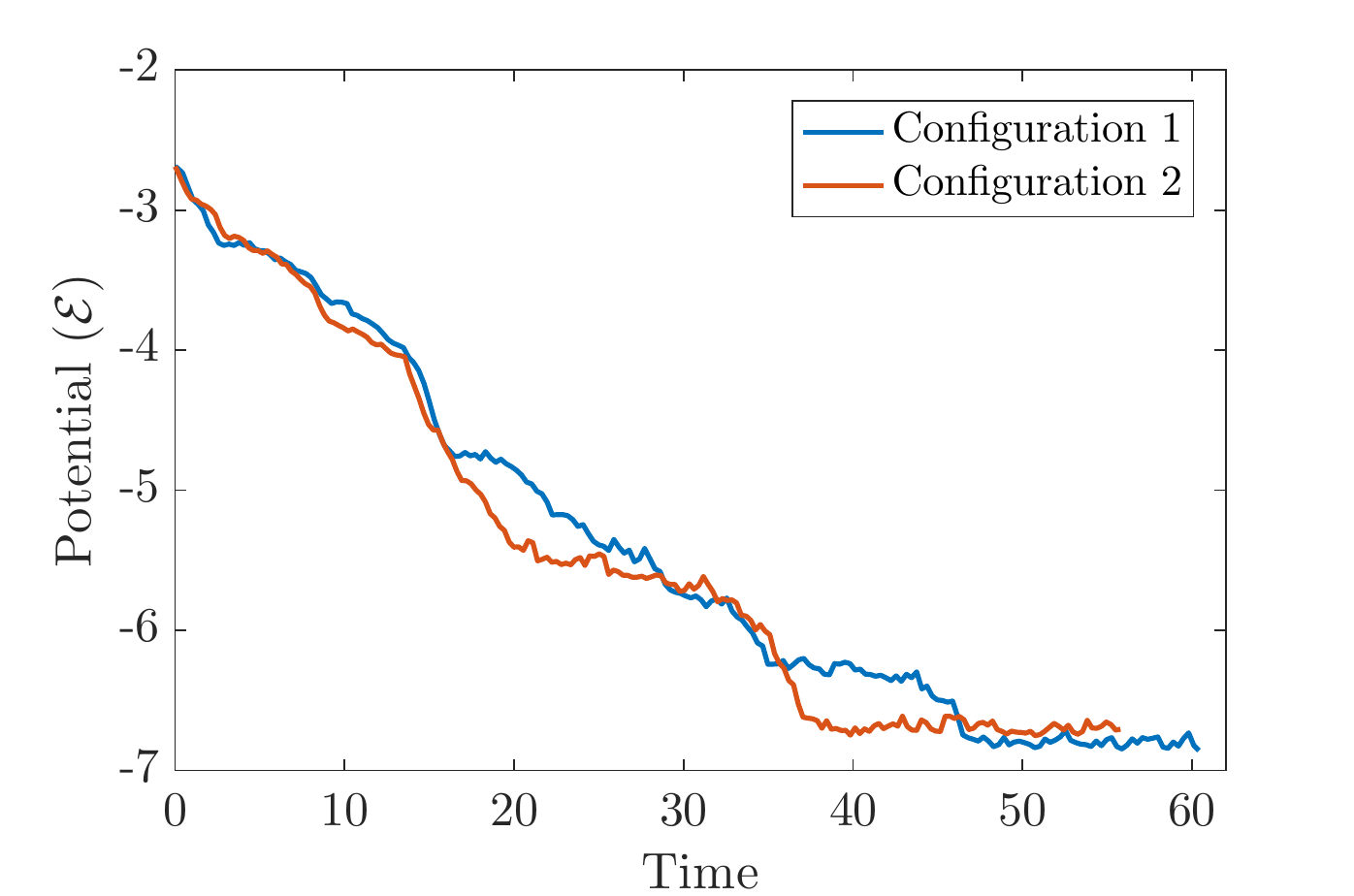}}
\end{tabular}
\caption{Formation of different clusters from the same initial positions of particles in the presence of noise of strength $\sigma^2 = 0.01$. Variation in potential of system with time during clustering for the cases shown.}
\label{fig:potcompare}
\end{center}
\end{figure}

\textit{Low noise} conditions are characterised by $\sigma^2\ll\Delta \mathcal{E}^{\scriptsize \mbox{barr}}$, which is not sufficient to change the configuration of a stable cluster once formed. However, it can influence the clustering dynamics by rearranging the particles as the system moves down the steepest gradient in interaction potential, as seen in Figure~\ref{fig:potcompare}. 

\begin{figure}[t]
\begin{center}

\includegraphics[height=8.5cm]{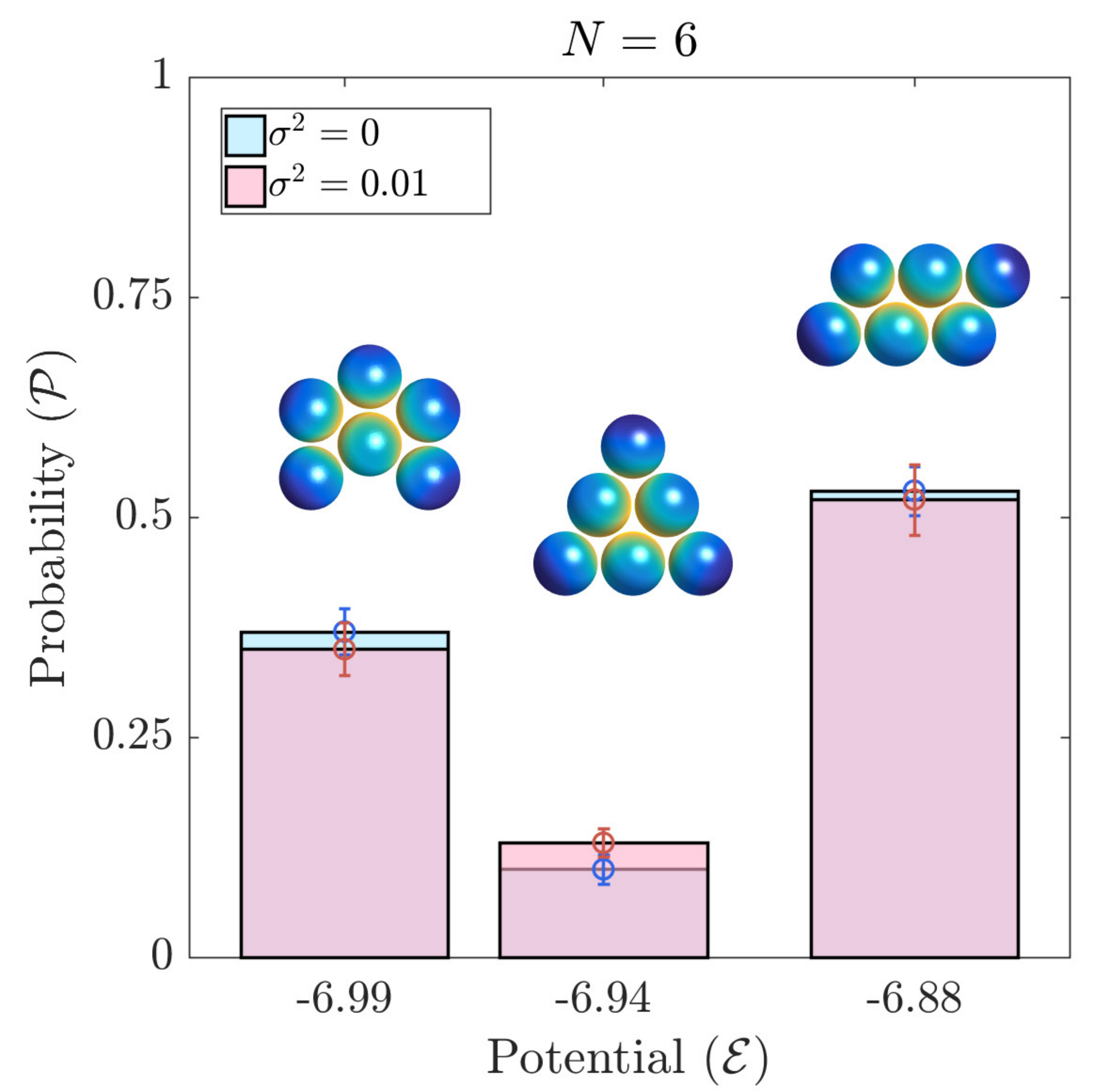}
\caption{Comparison between probabilities of formation of 6 particle cluster configurations in the absence of noise and with low noise shows no significant difference in the statistics. The clustering was performed with $R_{max}=16$ in both cases. In the case of low noise ($\sigma^2=0.025$), probabilities are computed from 200 random initial arrangements of particles with 100 trials per arrangement.}
\label{fig:low_stats}
\end{center}
\end{figure}

Its influence is particularly important when particles are far apart, i.e. when differences in $\mathcal{E}$ are small and of order $\sigma^2$. For this reason, one may wrongly presume that noise would change the probability statistics for the formation of cluster configurations. Instead, the probability statistics remain identical to that in the absence of noise (see Figure~\ref{fig:low_stats}).

 This result can be explained as follows: although a low background noise is expected to enable the system to explore neighboring routes in the configuration space to minimize the system's potential, $\mathcal{E}$, during the clustering phase, it is only effective in the initial stages where system has low $\mathcal{E}$; in the later stages, the magnitude of the noise becomes too small in front of the deterministic velocity (determined by $\grad_{\Rb_i}\mathcal{E}$) to significantly alter the particles' trajectories. However, a lower potential during the initial stages does not ensure a low potential of the cluster thus formed (as observed in Figure~\ref{fig:potcompare}). Thus, the noise just effectively redistributes the initial arrangement of the particles. But this randomness is already taken into account in the probability of formation of different clusters by considering a large number of random initial positions. As a result the collective propulsion statistics in Figure~\ref{meanmax} remain unmodified in low-noise conditions. 

\subsection{High noise}
\begin{figure}[h]
\begin{center}
\begin{tabular}{lc}
\includegraphics[height=7cm]{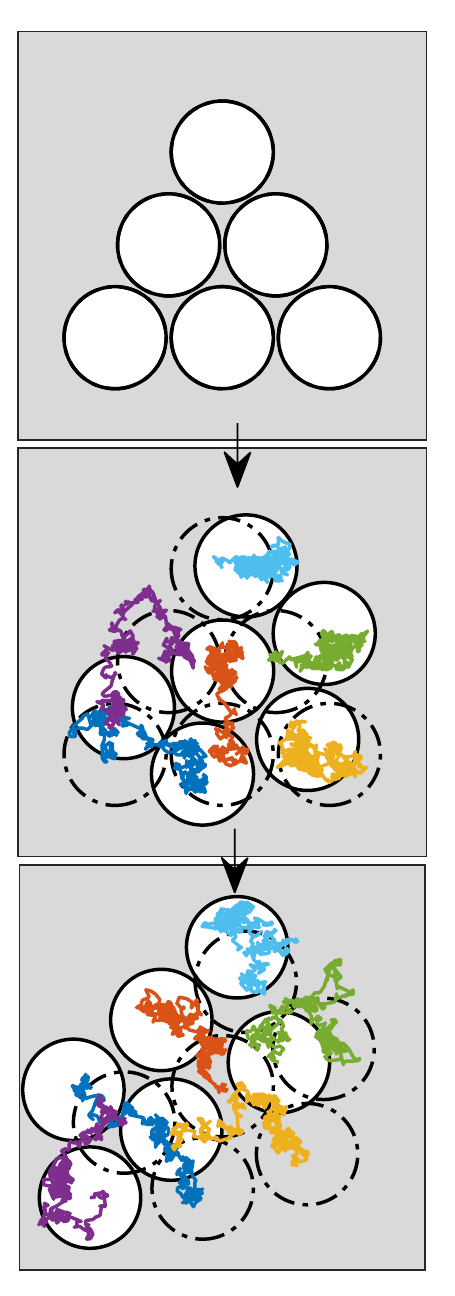} &
\includegraphics[height=7cm]{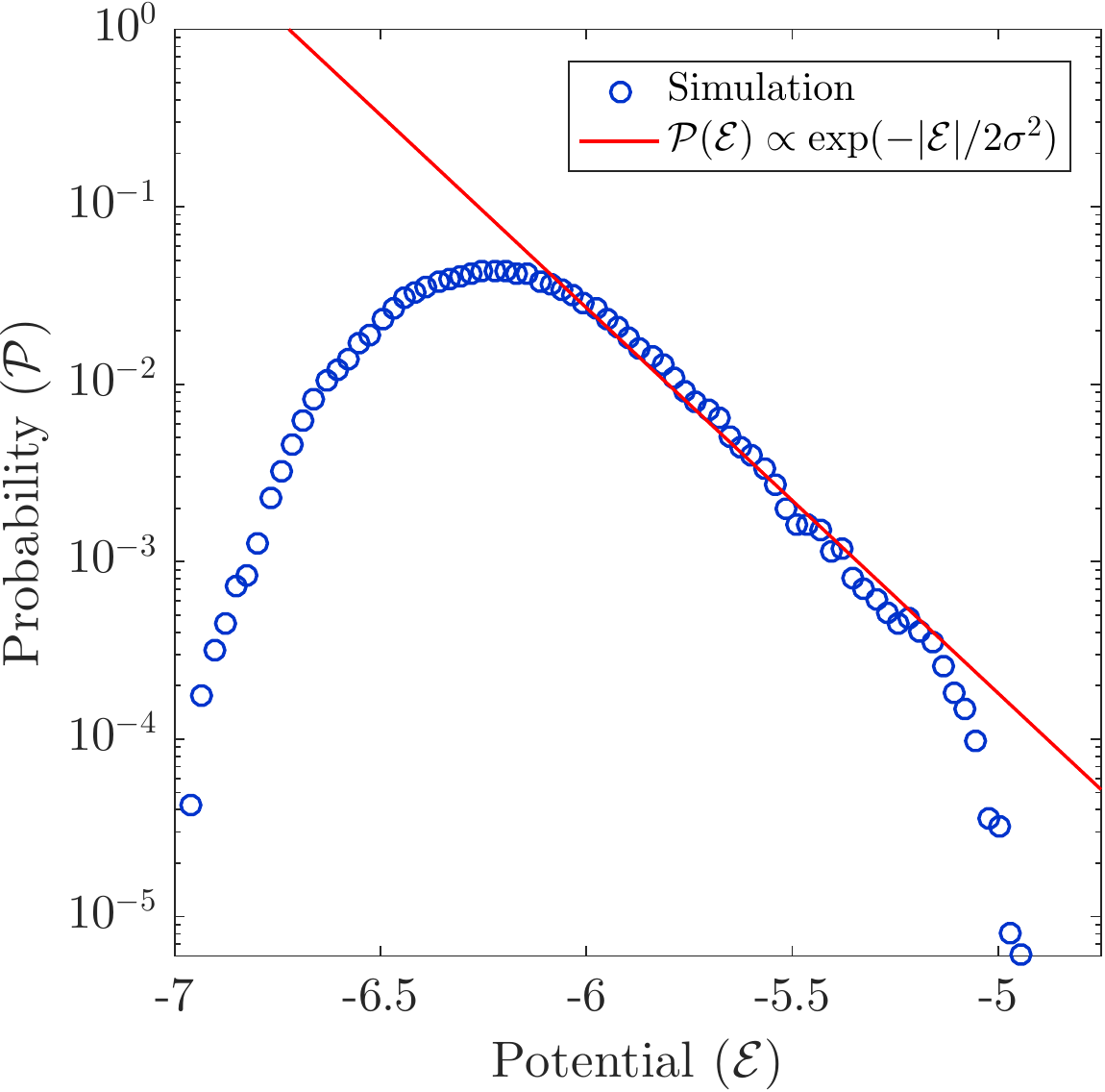}
\end{tabular}
\caption{Formation of different cluster configurations by rearrangement of particles over time due to high noise, $\sigma^2 = 0.1$. Also shown is the probability distribution of various cluster configurations which follow a Boltzmann distribution. Since the average separation between particles are larger due to the continuous high noise, $\mathcal{E}$ takes larger values.}
\label{boltz}
\end{center}
\end{figure}

In the presence of sufficiently large noise, $\sigma^2 ~\sim O(\Delta \mathcal{E}_{p \to q}^{\scriptsize \mbox{barr}})$, a cluster, once formed, continuously transitions from one configuration to another without dislocating fully since the noise intensity remains much smaller than $|\mathcal{E}_q|$. This situation is in stark contrast with the low-noise dynamics for two main reasons: (i) it is not possible anymore to define a fixed cluster shape to the arrangement of the particles which continuously evolves in time and transitions between all the available configurations, and (ii) the clustering dynamics does not play a significant role anymore. Indeed, since the particles' arrangement can be reconfigured, which cluster shape was reached in the first place is not relevant. Instead, how much time the particles spend in a particular configuration is now the significant information. This can be characterized as the probability to obtain a given configuration at any time, which is the result of the equilibrium between the background noise and the phoretic attraction. Not surprisingly, this probability now follows a Boltzmann distribution (Figure~\ref{boltz}). However, since multiple intermediate configurations (with the same potential) exist, the probability of a particular stable configuration cannot be defined.

\section{Conclusions}
\label{sec:conclusions}
The results presented in this work therefore establish a fourth route to self-propulsion of active colloids. Unlike previously-identified strategies which relied either on an asymmetric design of the particle or the non-linear convective transport of solutes by the phoretic flows, self-propulsion of individually non-motile yet active particles is achieved here by non-symmetric interactions between multiple particles. Under the effect of attractive phoretic attractions and steric constraints, particles form geometrically-asymmetric clusters that are able to maintain the asymmetric concentration fields required for propulsion. The self-propulsion velocities are much lower (by at least an order of magnitude) than typical relative velocities between particles during particle clustering, but lead nonetheless to a net migration of the particles.

Even though the governing dynamics of the particles (in the absence of external noise) are completely deterministic, multiple particle arrangements can be reached depending on the detailed dynamics of the cluster formation and balance between phoretic attraction and steric repulsion. Slightly different initial positions of the different particles may therefore lead in fundamentally different propulsion characteristics, thereby introducing an inherent stochasticity in the system. Focusing for simplicity on two-dimensional arrangements of the particles (the hydrodynamics and solute diffusion are three dimensional), the probability of formation of the different cluster shapes was determined for increasing $N$ using Monte-Carlo simulations of the phoretic clustering for $N$ particles with initially-random positions. These simulations were performed using a reduced-order model of the clustering process obtained by superimposing the velocity induced on a given particle by each of its neighbors as in a two-particle system, for which an analytical solution was first obtained. 

Using these probabilities and the velocity of the different clusters computed numerically using a regularized Boundary Element Method, the statistics of the collective self-propulsion were obtained, namely the mean, maximum and most probable velocities, for both translation and rotation.  The maximum attainable velocity was found to increase with $N$: for larger clusters, a greater degree of asymmetry can be achieved leading to larger velocities.  The mean propulsion velocity on the other hand was found to saturate for increasing $N$: the most eccentric clusters with large velocity also become less and less probable as $N$ increases, while the most probable clusters exhibit a more compact geometry with velocities close to the mean value. 

The particles are continuously in a state of non-equilibrium. The lack of external noise to relax the system creates probability statistics of various configurations which are strikingly different from equilibrium statistics. This difference directly affects the velocity statistics and we observe a larger mean velocity compared to that expected from a purely equilibrium process. Yet thermal noise and Brownian motion can become important for smaller particles, and their impact on the present result was tested by introducing a Brownian component in the motion of the active particles in addition to their deterministic phoretic clustering. Two different situations are observed depending on the magnitude of the noise with respect to the phoretic forces maintaining the cohesion of the formed clusters. The cluster probability and velocity statistics are found remarkably robust to low noise amplitude (i.e. thermal fluctuations that are unable to break or reconfigure a given cluster): such noise levels essentially redistribute the already random position of the particles in their initial state (before clustering).  In contrast, higher levels of thermal noise can lead to cluster break-up or reconfiguration. In the latter, a given cluster undergoes continuous transition between different configurations, and the probability to find a given shape  follows an equilibrium probability distribution. In that case, the group of $N$ particles continuously reconfigure leading to changes in its translational and rotational velocity. This provides it with a self-propelled motion which is reminiscent in some regards to the run-and-tumble behaviour of bacteria where self-propulsion in a given configuration is followed by a geometric reorganization that modifies its swimming direction and velocity magnitude~\citep{wykes16}. Although the process is quite different here, in particular because the relative length of a run is significantly smaller than for bacteria, it is expected to enhance transport by diffusion. 
 
The central goal of the present work is to characterize the emergence of the self-propulsion from phoretic and hydrodynamic interactions of individually-non-motile isotropic particles. We demonstrated this property for the simpler configuration of two-dimensional motion of the particles. Yet, the different tools developed both analytically and numerically directly apply to three-dimensional and the present results could easily be extended to generic 3D clusters and associated propulsion properties, the main difficulty residing in the classification of the different cluster shape. We finally remark that the formalism developed in this manuscript, both numerical and analytical, could also  be extended to address the collective dynamics of mixtures of particles of different chemical properties (e.g. mixture of active and inert, or isotropic and Janus particles \cite{soto14,schmidt2018}).

\acknowledgements
This project has received funding from the European Research Council (ERC) under the European Union's Horizon 2020 research and innovation programme (grant agreement No 714027 to SM) and the Engineering and Physical Sciences Research Council of the United Kingdom (grant EP/R041555/1 to TDMJ). The authors would like to acknowledge valuable discussions with Eric Lauga.

\appendix

\section{Effect of a finite regularization parameter}\label{app:reg}

We present briefly here the derivation of Eq.~\eqref{eq:regeffect} which provides the leading order correction introduced by a finite regularization parameter $\epsilon$ in the regularized Boundary Element Method framework. The general goal in BEM is to express the following integral
\begin{equation}
I(\xb_0)=\int_{V_f}c(\xb)\phi^\epsilon(\xb,\xb_0)\mathrm{d}\mathcal{S}_\xb,
\end{equation}
with
\begin{equation}
\quad \phi^\epsilon(\xb,\xb_0)=\frac{15\epsilon^4}{8\pi(|\xb-\xb_0|^2+\epsilon^2)^{7/2}},
\end{equation}
in terms of the local value of $c(\xb_0)$ and its gradients, when $\xb_0$ is on the surface of the particle. Here, $V_f$ is the fluid volume outside the particle. To perform this computation, we consider a local reference frame centred on $\xb_0$ and such that $\nb(\xb_0)=\eb_z$. The fluid domain outside the particle is locally defined as $z+(\kappa_xx^2+\kappa_yy^2)/2\geq 0$ with $\kappa_x$ and $\kappa_y$ are the two principal curvatures (this implicitly assumes that $|\kappa|\epsilon\ll 1$ with $\kappa$ the mean curvature). 

First, $c(\xb)$ is expanded in Taylor series around $\xb_0$:
\begin{equation}
I(\xb_0)=c(\xb_0)\int_{V_f}\phi^\epsilon(\xb,\xb_0)\mathrm{d}\mathcal{S}_\xb+\nabla c(\xb_0)\int_{V_f}\phi^\epsilon(\xb,\xb_0)\xb\mathrm{d}\mathcal{S}_\xb+\hdots
\end{equation}
and using a local spherical polar coordinate system with $\mu=\cos\theta$ and $\mu_c(r,\phi)=-(\kappa_x\cos^2\phi+\kappa_y\sin^2\phi)r/2$ the parametric description of the local particle surface
\begin{align}
I(\xb_0)=&\,\frac{15\epsilon^4c(\xb_0)}{8\pi}\int_0^{2\pi}\int_0^\infty\int_{\mu_c(r,\phi)}^1\frac{r^2\dd r\dd\mu\dd\phi}{(r^2+\epsilon^2)^{7/2}} +\frac{15\epsilon^4}{8\pi}\nabla c(\xb_0)\cdot\int_0^{2\pi}\int_0^\infty\int_{\mu_c(r,\phi)}^1\frac{r^2\mathbf{r}\dd r\dd\mu\dd\phi}{(r^2+\epsilon^2)^{7/2}}+\hdots\nonumber\\
=&\,\frac{15\epsilon^4c(\xb_0)}{4}\int_0^\infty\frac{r^2\dd r}{(r^2+\epsilon^2)^{7/2}}\left(1+\frac{\kappa r}{2}\right)+\frac{15\epsilon^4}{8}(\nb(\xb_0)\cdot\nabla c(\xb_0))\int_0^\infty\frac{r^3\dd r}{(r^2+\epsilon^2)^{7/2}}+\hdots\nonumber\\
=&\,c(\xb_0)\left(\frac{1}{2}+\frac{\kappa\epsilon}{4}\right)+\frac{\epsilon\nb(\xb_0)\cdot\nabla c(\xb_0)}{4}+\hdots
\end{align}
The development above includes the first two leading order in $\epsilon$: the $O(\epsilon^0)$ contribution is the classical result for a singular Boundary Integral Methods and the $O(\epsilon)$ accounts for the finite size of the regularization in comparison with (i) the local curvature of the surface and (ii) the typical length scale for the variations of $c$.

\section{Method of reflections} \label{app:mor}
The method of reflections is a classical technique to construct an asymptotic solution of Laplace or Stokes problem around several or many bodies using an iterative approach~\cite{kimkarrila}. This asymptotic development is made in terms of $\varepsilon=1/d$, i.e. the ratio of the particles' radius to typical distance among them. Taking the concentration problem as an example, the solution $c_j^0$ is first obtained for each particle $j$ in the absence of all the other particles. These solutions are then corrected at each stage (i.e. each reflection) near each particle $k\neq j$ to cancel the extra contribution to the boundary conditions introduced at the surface of this particle by the previous reflections.

\subsection{Diffusion problem}
\label{sec:mor_laplace}
The Laplace problem for the concentration around all the particles of unit radii can be formulated in full as follows
\begin{align}
\nabla^2 c&=0,\\
\nb_j\cdot\nabla c&=-A\quad \textrm{for   } r_j=1,\\
c&\rightarrow 0\qquad \textrm{for   }r\rightarrow\infty.
\end{align} 
As for the main text of the manuscript, $\rb_j=\rb-\Rb_j$ is the position relative to particle $j$, $\nb_j$ its outward-pointing normal, $d_{jk}=|\Rb_j-\Rb_k|$ the relative distance of particles $j$ and $k$, and $\eb_{jk}=(\Rb_k-\Rb_j)/d_{jk}$ is the unit vector point from particle $j$ to particle $k$.

The solution of this problem for an isolated particle $l$  is simply $c_l^0=A/r_l$. Expanding this solution near particle $k\neq l$ (i.e. for small $r_k$), 
\begin{equation}
c^0_l = \frac{A}{r_l}=\frac{A}{d_{kl}} \sum_{p=0}^{\infty} \left(\frac{r_k}{d_{kl}}\right)^p \; L_p\left(\frac{\rb_k\cdot\mathbf{e}_{kl}}{r_k}\right),
\label{c0k}
\end{equation}
where $L_p(\mu)$ is the Legendre polynomial of degree $p$.

The first reflection solution $c_k^1$ therefore satisfies $\nabla^2 c_k^1=0$, decays at infinity, and at the surface of particle $k$ 
\begin{align}
\nb_k\cdot\nabla c_k^1&=-\nb_k\cdot \sum_{\substack{l=1\\ l\neq k}}^N\nabla c_l^0\quad \textrm{for   } r_k=1.
\end{align} 
The unique solution is obtained as
\begin{equation}
c_k^1 = A \sum_{\substack{l=1\\l\neq k}}^N\sum_{p=1}^{\infty}\frac{p}{p+1} \left(\frac{1}{d_{kl}}\right)^{p+1} \left(\frac{1}{r_{k}}\right)^{p+1} L_p\left(\frac{\rb_k\cdot \mathbf{e}_{kl}}{r_k}\right).
\label{c1j}
\end{equation} 
Note that this expansion includes no source term ($p=0$) since $c_l^0$ satisfies $\nabla^2 c_l^0=0$ in the vicinity of particle $k$.

The leading order correction introduced by the  second reflection on particle $j$ is a response to the uniform gradient $\mathbf{G}^1_j$, created near particle $j$ by the source dipole contribution in each $c_k^1$ ($k\neq j$) in the above equation, namely
\begin{equation}
\mathbf{G}^1_j=\sum_{\substack{k=1\\k\neq j}}^N\nabla c^1_k|_{r_j=0}
=\frac{A}{2} \sum_{\substack{k,l=1\\k\neq (j,l)}}^N\left(\frac{1}{d_{kl}}\right)^2\left[ \frac{ \mathbf{e}_{kl}}{r_k^3}-\frac{3(\rb_k\cdot\eb_{kl})\rb_k}{r_k^5}\right].
\end{equation}
and the leading order contribution to the second reflection writes
\begin{align}
c^2_j & = \frac{\mathbf{G}_j^1 \cdot \mathbf{n}_j}{2r^2_j}= \frac{A}{4}\sum_{\substack{k,l=1\\k\neq (j,l)}}^N \frac{(\mathbf{e}_{kl}-3(\mathbf{e}_{kl}\cdot \mathbf{e}_{jk})\mathbf{e}_{jk})}{d^3_{jk} d^2_{kl} r^2_j} \cdot \mathbf{n}_j.
\label{c2j}
\end{align}
Using these results, the concentration $c_j^s$ and slip velocity $\tilde\ub_j=M(\mathbf{I}-\nb_j\nb_j)\cdot\grad c_j^s$ at the surface of particle $j$ using two reflections, are given by
\begin{align}
c^s_j &= C^j_0 + \mathbf{C^j_1} \cdot \nb_j +\mathbf{C}^j_2 :(\nb_j\nb_j)+ ...,\\
\tilde{\mathbf{u}}_j &=  \mathbf{C^j_1}+\mathbf{C_2^j}\cdot\nb_j+...,
\end{align}
with
\begin{align}
\mathbf{C_1^j} & =A\left[  \sum_{\substack{k=1\\k\neq j}}^N \frac{3\,\eb_{jk}}{2d^2_{jk}}  + \sum_{\substack{k,l=1\\k\neq (j,l)}}^N \frac{(\mathbf{I}-3\mathbf{\hat{e}}_{jk}\mathbf{\hat{e}}_{jk}) \cdot \mathbf{\hat{e}}_{kl}}{4d^3_{jk}d^2_{kl}}  +  O\left(\varepsilon^6\right)\right],\\
\mathbf{C^j_2}&=A\left[\frac{5}{6}\sum_{\substack{k=1\\k\neq j}}^N\frac{3\,\eb_{jk}\eb_{jk}-\mathbf{I}}{d_{jk}^3}+O(\varepsilon^6)\right].
\label{C1}
\end{align}

\subsection{Hydrodynamic problem for free particles} \label{app:free}
We now turn to the iterative solution of the Stokes problem for individually force-free particles, namely
\begin{align}
\eta\nabla^2\ub=\grad p,&\qquad \nabla\cdot\ub=0,\qquad \ub(\infty)=0\\
\nonumber\\
\forall j, \quad \ub=\Ub_j+&\Omegab_j\times\rb_j+\tilde\ub_j\quad\textrm{ on   }r_j=1,\\
\forall j, \quad \int_{r_j=1}\sigmab\cdot\nb&\,\dd S=\int_{r_j=1}\rb_j\times(\sigmab\cdot\nb)\,\dd S=0.
\end{align} 

Following the general framework of the method of reflections, the swimming problem for a single isolated particle $k$ is solved for the velocity, rotation rate and stresslet of this particle as 
\begin{align}
\mathbf{U}^0_k & = -\langle\tilde\ub_k\rangle  = -\frac{2M}{3}\mathbf{C_1^k},\\
\Omegab_k^0&=-\frac{3}{2}\langle \nb_k\times\tilde\ub_k\rangle=0,\\
\mathbf{S}_k^0&=-10\pi\langle \nb_k\tilde\ub_k+\tilde\ub_k\nb_k\rangle=-8\pi M\mathbf{C_2^k},
\label{fU0}
\end{align}

The leading order flow field generated by particle $j$ is then
\begin{equation}
\ub_k^0=-M\left(\frac{\rb_k\rb_k}{r_k^5}-\frac{\mathbf{I}}{3r_k^3}\right)\cdot\mathbf{C_1^k}+3M\left(\frac{\rb_k\rb_k\rb_k}{r_k^5}\right):\mathbf{C_2^k}.
\end{equation}
The development above retains only the source and force dipoles as fundamental singularities. Force quadrupoles, which have the same far-field decay as that of the source dipole ($1/r^3$), are nevertheless neglected because their intensity is determined by $\mathbf{C_3^j}$ which is of subdominant order. The result of the first reflection is the swimming and rotation velocities of force- and torque-free particles in an ambient flow field, and are determined directly using Faxen's laws for spherical particles. As a result,
 \begin{align}
\Ub_j^1&=-M\sum_{\substack{k=1\\ k\neq j}}^N\left[\frac{1}{3d_{jk}^3}(3\eb_{jk}\eb_{jk}-\mathbf{I})\cdot\mathbf{C_1^k}+\frac{3}{d_{jk}^2}(\eb_{jk}\cdot\mathbf{C_2^k}\cdot\eb_{jk})\eb_{jk}\right],\\
\Omegab_j^1&=3M\sum_{\substack{k=1\\ k\neq j}}^N\frac{(\mathbf{C_2^j}\cdot\eb_{jk})\times\eb_{jk}}{d_{jk}^3}\cdot
\end{align}
Using the results for the Laplace problem (Section~\ref{sec:mor_laplace}), the velocity of force-free particle $j$ is obtained as
\begin{align}
\Ub_j=AM\left[-\sum_{\substack{k=1\\ k\neq j}}^N\frac{\eb_{jk}}{d_{jk}^2}
+\frac{5}{2}\sum_{\substack{k,l=1\\ k\neq j\\l\neq k}}^N\frac{\left(3(\eb_{jk}\cdot\eb_{kl})^2-1\right)\eb_{kj}}{d_{jk}^2d_{kl}^3}+O\left(\varepsilon^6\right)\right],
\end{align}
and the velocity of the center of mass, $\displaystyle\Ub_{CM}=\frac{1}{N}\sum_{j=1}^N\Ub_j$, is
\begin{equation}
\Ub_{CM}=\frac{5AM}{2N}\left[\sum_{\substack{j,k,l=1\\ k\neq j\\ l\neq k}}^N\frac{\left(3(\eb_{jk}\cdot\eb_{kl})^2-1\right)\eb_{kj}}{d_{jk}^2d_{kl}^3}+O(\varepsilon^6)\right].
\end{equation}

For free (i.e. non-touching) particles, individual particles' velocities scale as $1/d^2$ while their center of mass moves with an $1/d^5$ velocity.

\subsection{Hydrodynamic problem for rigid clusters} \label{app:cluster}
When the particles are rigidly-bound in a cluster, they are not individually force-free, but, the total force and torque on the cluster are zero. Assuming that the non-hydrodynamic interaction forces between the particles are central, the net torque on particle $j$ around its center of mass must vanish. Hence,
\begin{equation}
\forall j,\quad \int_{r_j=1} \sigmab \cdot \nb_j=-\mathbf{F}_j, \;\;\; \int_{r_j=1} \rb_j \times (\sigmab \cdot \nb_j) \;\;\text{d}S =0.
\end{equation}

Within the framework of the method of reflections, solving the hydrodynamic problem for particle $k$, leads to
\begin{align}
\mathbf{U}^0_k & = -\langle\tilde{\mathbf{u}}_k\rangle + \frac{\mathbf{F}_k}{6 \pi} = -\frac{2M}{3}\mathbf{C_1^k} + \frac{\mathbf{F}_k}{6 \pi}, \label{U0} \\
\Omegab^0_k & = -\frac{3}{2}\langle\mathbf{n}_j \times \tilde{\mathbf{u}}_j\rangle = 0.
\end{align}
Particle $k$ is not force-free anymore, and its flow signature must now also include a contribution from an additional Stokeslet and source dipole associated with $\Fb_k$:
\begin{align}
\ub_k^0=  -M\left(\frac{\rb_k\rb_k}{r_k^5}-\frac{\mathbf{I}}{3r_k^3}\right)\cdot\mathbf{C_1^k}+3M\left(\frac{\rb_k\rb_k\rb_k}{r_k^5}\right):\mathbf{C_2^k} +\frac{\Fb_k}{8\pi}\left(\frac{\mathbf{I}}{r_k}+\frac{\rb_k\rb_k}{r_k^3}\right)+\frac{\Fb_k}{24\pi}\left(\frac{\mathbf{I}}{r_k^3}-\frac{3\rb_k\rb_k}{r_k^5}\right).
\end{align}
From Faxen's law, the correction to the velocity of particle $j$ introduced by the first reflection is therefore
\begin{align}
\Ub_j^1=  -M\sum_{\substack{k=1\\k\neq j}}^N\left[\frac{1}{3d_{jk}^3}(3\eb_{jk}\eb_{jk}-\mathbf{I})\cdot\mathbf{C_1^k}+\frac{3}{d_{jk}^2}(\eb_{jk}\cdot\mathbf{C_2^k}\cdot\eb_{jk})\eb_{jk}\right] +\sum_{\substack{k=1\\k\neq j}}^N\left[\frac{\Fb_k}{8\pi d_{jk}}(\mathbf{I}+\eb_{jk}\eb_{jk})+\frac{\Fb_k}{12\pi d_{jk}^3}(\mathbf{I}-3\eb_{jk}\eb_{jk})\right].
\end{align}
Using the result for the concentration problem and the rigid body kinematics, $\Ub_j=\Ub_{CM}+\Omegab_{CM}\times\Rb_j$,
\begin{equation}
\Ub_{CM} + \Omegab_{CM} \times \Rb_j - \mathbf{U}^\textrm{free}_j = \sum_{k} \mathbf{K}_{jk} \cdot \mathbf{F}_k, 
\end{equation}
with 
\begin{align}
\Ub_j^\textrm{free}&=-AM\left[\sum_{k\neq j}\frac{\eb_{jk}}{d_{jk}^2}+\frac{5}{2}\sum_{\substack{k\neq j\\l\neq k}}\frac{[3(\eb_{jk}\cdot\eb_{kl})^2-1]\eb_{jk}}{d_{jk}^2d_{kl}^3}\right],\\
\mathbf{K}_{jj}&=\frac{\mathbf{I}}{6\pi},\qquad 
\mathbf{K}_{jk}=\frac{\mathbf{I}+\eb_{jk}\eb_{jk}}{8\pi d_{jk}}-\frac{3\eb_{jk}\eb_{jk}-\mathbf{I}}{12\pi d_{jk}^3}\qquad \textrm{ for   }k\neq j.
\end{align}
The translation and rotation velocity of the cluster are then obtained by imposing that the cluster as a whole is force- and torque-free, i.e. $\displaystyle\sum_j\Fb_j=0,\;\sum_j \Rb_j\times\Fb_j=0$. The resulting leading order velocity of the center of mass, $\Ub_{CM}$, now scales as $1/d^3$.


\end{document}